\documentclass[twocolumn,superscriptaddress,nature]{revtex4-2}
\usepackage[colorlinks=true,citecolor=blue,linkcolor=blue,urlcolor=blue,]{hyperref}
\usepackage{graphicx, nicefrac, textcomp}
\usepackage[normalem]{ulem}
\usepackage{changes}
\usepackage{chngcntr}
\usepackage{amssymb}
\usepackage{amsmath}
\usepackage{soul}
\usepackage{color}
\usepackage{float}
\usepackage{textcomp}
\usepackage{amstext}
\usepackage{graphicx}
\usepackage{gensymb}
\usepackage{graphics}\usepackage{subfigure}\usepackage{longtable}\usepackage{pstricks}\usepackage{dcolumn}\usepackage{bm}
\usepackage{siunitx}
\usepackage{wasysym}

\begin{document}

 \title{Exerting chemical pressure on the kagome lattice as frustration control in the kapellasite family $A$Cu$_3$(OH)$_{6+x}$(Cl,Br)$_{3-x}$}
\author{Jonas Andreas Krieger}
\affiliation{PSI Center for Neutron and Muon Sciences CNM, CH-5232 Villigen, Switzerland}
\author{Thomas James Hicken}
\affiliation{PSI Center for Neutron and Muon Sciences CNM, CH-5232 Villigen, Switzerland}
\author{Hubertus Luetkens}
\affiliation{PSI Center for Neutron and Muon Sciences CNM, CH-5232 Villigen, Switzerland}
\author{Reinhard K. Kremer}
\affiliation{Max-Planck-Institute for Solid State Research, Heisenbergstra{\ss}e 1, 70569 Stuttgart, Germany}
 \author{Pascal Puphal}
\email[]{puphal@fkf.mpg.de}
\affiliation{Max-Planck-Institute for Solid State Research, Heisenbergstra{\ss}e 1, 70569 Stuttgart, Germany}
\affiliation{2nd Physics Institute, University of
Stuttgart, 70569 Stuttgart, Germany}

\date{\today}

\begin{abstract}
The kapellasite family $A$Cu$_3$(OH)$_{6+x}$(Cl,Br)$_{3-x}$ forms a series of compounds, wherein the chemical pressure realized by the $A-$site cation tunes the spin exchange in the frustrated distorted kagome lattice. Via hydrothermal synthesis we have grown single crystals of the whole series for $A=$ rare-earths and found  a clear structural transition to a superstructure variant at a specific chemical pressure level  exerted by $A$~=~Dy.  Phases with crystal structures in the vicinity of this superstructure transition realize a distorted kagome lattice of the Cu$^{2+}$ cations with characteristic features of a spin liquid. Here, subtle structural disorder as well as controlled chemical pressure can stabilize the spin liquid phase. We study the crystal structure and the magnetic ground state of these phases via single crystal x-ray diffraction, magnetic susceptibility, specific heat and muon spin spectroscopy measurements.
\end{abstract}

\maketitle

\begin{figure}[h]
\centering
\includegraphics[width=1\columnwidth]{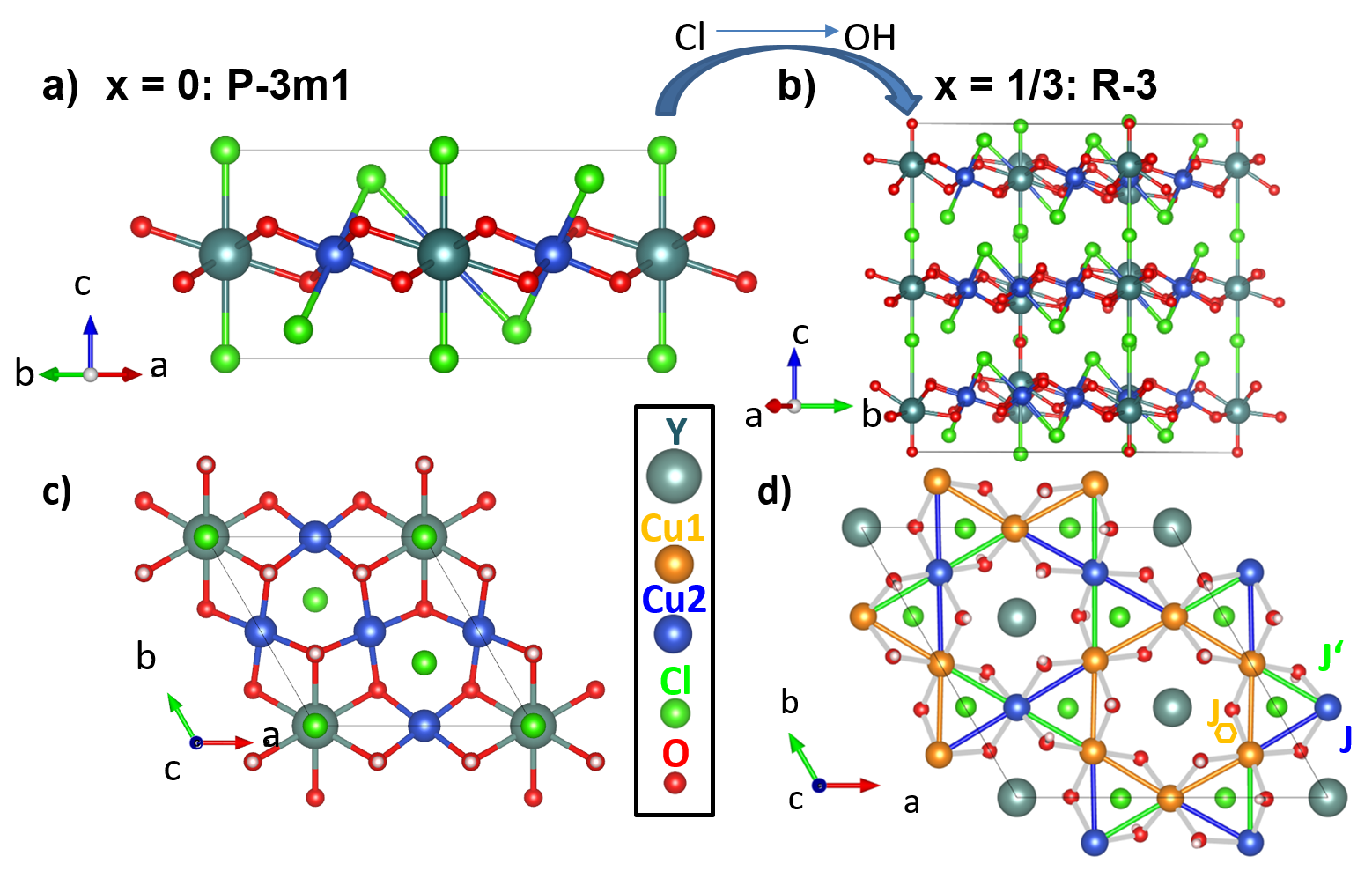}
\caption{\textbf{Crystal structure of $A$Cu$_3$(OH)$_{6+x}$(Cl,Br)$_{3-x}$. a,c)} for large $A-$ site cations $x=0$ realizing the parent $P\overline3m$1 structure. The structure is projected along (110) in a) and along (001) in c). \textbf{b,d)} the superstructure with space group $R\overline3$ for small $A-$site cations beyond Y leading to $x=1/3$. The crystal structure is projected along (110) in b) and  along (001) in d). As displayed in the legend: the $A-$atoms are displayed in light gray, the Cu-atoms in blue/orange, the O-atoms in red-white color and the Cl-atoms in green. For better visibility the hydrogen atoms in the hydroxyl groups have been omitted for in-plane view. The magnetic exchange paths are shown as color coded bonds in d) with the hexagon $J_h$ orange line, $J$ as blue line and the short bonded $J'$ as green line.
}
\label{struc}
\end{figure}

Antiferromagnetic kagome layers, are prime candidates to search for an experimental realization of the quantum spin-liquid (QSL) phase \cite{Balents10,Chamorro2020}, characterized by the absence of static magnetic order down to lowest temperatures with macroscopic entanglement and fractional excitations. 
Cu-based systems with kagome layers of Cu$^{2+}$ spin $S$~=~1/2 cations are considered ideal QSL candidates, with one of the first experimental QSL phases observed in herbertsmithite with composition ZnCu$_3$(OH)$_6$Cl$_2$ \cite{mendels07,Han2012}. Recently, the disordered case of a Dirac spin liquid (SL) \cite{Jeon2024,Zeng2024} is found within a related new family of compounds with composition $A^{n+}$Cu$_3$(OH)$_6X_n$  wherein $X$ can be F, Cl, Br, NO$_3$, SO$_4$ and $A$ both a divalent or trivalent cation as summarized in Ref.~\cite{Puphal2018}.  The majority of these systems crystallize with kapellasite-related structures, a structural polymorph of the herbertsmithite structure with AA-stacked perfect kagome planes \cite{Boldrin2016} (see Figure \ref{struc}). For trivalent cations $A^{3+}$, e.g. $A$~=~ Nd~-~Dy, Y,  an additional halogen atom is included, resulting
in compounds with
composition $A^{3+}$Cu$_3$(OH)$_6$Cl$_3$ crystallizing with a trigonal crystal structure (space group $P\overline{3}m$1, no. 164) \cite{Puphal2018}.
For $A~=~$ Nd, Sm  canted antiferromagnetic order of the Cu magnetic moments has been found at around 20~K and 18~K \cite{Sun2017}. Polarization of the rare-earth moments is indicated by a second anomaly in the magnetic susceptibility at 8~K and 6~K. For $A$~=~Eu, we have reported two magnetic transitions at $T_{\rm N}$~=~14.7~K and $T_{\rm N}$~=~8.7~K, where the magnetism shows enhanced anisotropy and increased effective magnetic moments of $\mu^{\perp}_{eff}=~2.5\mu_{B}$ and $\mu^{\parallel}_{eff}=3.25\mu_{B}$ at 350~K. \cite{Puphal2018} The enhancement over the Cu$^{2+}$ spin-only effective moment of 1.73~$\mu_{B}$  can be ascribed  to polarization from the Van Vleck excited moment of Eu$^{3+}$ cations. Also for $A$~=~Eu a magnetic long-range order (LRO) transition occurs with a sharp magnon peak appearing in the $E_g$ magnetic Raman continuum below the magnetic transition temperature of $T_{\rm N}$~=~17~K \cite{Fu21}.  For $A$~=~Gd, Tb, Dy similar magnetic transitions are observed at $T_{\rm N}$~=~17~K, 16~K, and 17~K, respectively \cite{Fu2021}. All these compounds show a similar additional magnetic transition from the polarization of the rare-earth ion. Gd, Tb, and Dy  show additional peaks in the heat capacity at 4.5~K, 7.5~K, and at 6~K, respectively \cite{Fu2021}. 
For $A$~=~Y, a magnetic transition at $T_{\rm N}$~=~15~K 
was detected by neutron diffraction \cite{Zorko2019} with clear magnetic Bragg reflections, but the Cu ordered magnetic moments are suppressed to 0.42(2)$\mu_B$. LRO  for $A$~=~Y is furthermore supported by static moments seen in $\mu$SR experiments \cite{Barthelemy2019,Zorko2019a}. The $q=0$ magnetic structure is similar to those of the   Cd- and the Ca-Kapellasite systems \cite{Okuma2017,Yoshida2017}. These two phases represent two examples of kapellasite family members with $A$ being a divalent ion, where the magnetic order  at 4~K and 7.2~K, respectively is ascribed to strong Dzyaloshinskii-Moriya (DM) interaction. \cite{Barthelemy2019,Zorko2019,Zorko2019a,Prelovsek2021} Hence, these $P\overline{3}m$1 structural variants constitute a family of the kapellasite-type compounds, where in all cases  magnetic field induced DM interaction causes ordering of the Cu ions with a transition for the rare-earth moments at around half $T_{\rm N}$, depending on by the paramagnetic rare-earth moments. 

Interestingly, for the nonmagnetic cation A~=~Y   the compound is not stable against water, leading to Cl~-~OH substitution. As a result phases with composition YCu$_3$(OH)$_{6+x}$Cl$_{3-x}$ are formed, until the end-member Y$_3$Cu$_9$(OH)$_{19}$Cl$_{8}$ for $x$~=~1/3 is reached, which  crystallizes with a trigonal structure (space group $R\overline{3}$, no. 148) \cite{Puphal2017,Kremer2024}. This substitutional variant shows a distortion of the  Cu lattice. The  Y atoms move out of the kagome plane, being pulled away by the shorter (OH)$^{-1}$ bond, thus generating  two non-equivalent Cu sites and a unique magnetic lattice with three different nearest-neighbor exchange interactions is required \cite{Puphal2017,Hering2022} to describe the experimental data. Similarly, Cd- and Ca-kapellasite have a H$_2$O molecule occupying the same interplane site \cite{Okuma2017,Yoshida2017} and Ca-kapellasite exhibits also Ca positions pushed out of the kagome planes, hence formation of a superstructure may also be expected \cite{Yoshida2017}.
For Y$_3$Cu$_9$(OH)$_{19}$Cl$_{8}$ chemical disorder-free crystals can be obtained which undergo LRO below 2.2~K, as spin waves were seen in inelastic neutron scattering, and fast damped oscillations in $\mu$SR were detected \cite{Chatterjee2023}. Multimagnon modes were detected by terahertz time-domain spectroscopy, further confirming LRO, and a 1/5 magnetization plateau was observed \cite{Biesner2022}. Finally, in  \textsuperscript{1}H-NMR a clear peak of the spin-lattice relaxation rate  $T^{-1}$ was detected at $T_{\rm N}$ \cite{Wang2023}.

 Recently, a phase with the very small trivalent cation  In$^{3+}$ was reported, crystallizing in space group $P\overline{3}m$1, however with two different Cu sites leading to highly distorted kagome layers \cite{Kato2024}. The compound shows short range order around 6~K and LRO around 2~K. Rare-earth ions with smaller ionic radii have not been studied yet.

In addition to the chlorides, the sister compound series YCu$_3$(OH)$_{6}$Br$_{2}$[Br$_x$(OH)$_{1-x}$] was also synthesized \cite{Chen2020}. Introducing additional site disorder by mixing Cl and Br the next level of complexity was reached \cite{Li2024,Shivaram2024,Xu2024}.   For pure bromides, the crystal structure determination proposed a mixture of the two end members of the series, similar to the Cl system, with partial occupancy of both Y sites, located within and out of the kagome layers.\cite{Sun2021}.  Variations of $x$ have been reported \cite{Zeng2022,Liu2022,Lu2022} even reaching the composition YCu$_3$(OH)$_{6.5}$Cl$_{2.5}$ \cite{Chen2020,Hong2022,Lu2022,Suetsugu2024} . This finding comes as a surprise since it results in $x$~=~1/2, thus exceeding the superstructure scenario of the compounds with $x=1/3$, i.e. a composition Y$_3$Cu$_9$(OH)$_{19}$Br$_8=$YCu$_3$(OH)$_{6.33}$Br$_{2.66}$ \cite{Lu2022}.  Unfortunately in most studies reported so far,  full crystal structure determinations were not performed. We recently were able to demonstrate  that clean crystals of $x=1/3$ again realize the superstructure \cite{Kremer2024}. 

The Br system attracted particular attention. A $T^2$ term in  the specific heat and a Dirac cone-shaped excitation spectrum found by inelastic neutron scattering experiments were interpreted as signatures of a Dirac spin-liquid ground state \cite{Liu2022,Zeng2022,Zeng2024,Han2024}. Thermal transport studies revealed thermally activated phonon-spin scattering indicating a gapped magnetic excitation spectrum \cite{Hong2022} and 1/9 and 1/3 magnetization plateaus were found \cite{Zheng2023,Suetsugu2024,Jeon2024}.   In addition, quantum oscillations in torque measurements were observed for the first time in an insulator \cite{Zheng2023}. 
However, $\mu$SR and NQR measurements revealed an inhomogeneous magnetic ground state with both fast and slow relaxing channels \cite{Lee2024}. 

As mentioned above smaller trivalent $A-$site rare-earth cations have not been reported, so far. Here, we report on the whole series of rare-earth elements $A$Cu$_3$(OH)$_{6+x}$(Cl,Br)$_{3-x}$ for both halogen anions, Cl and Br. In particular, focus on the syntheses,  structure determination by single crystal XRD, magnetic susceptibility, specific heat and $\mu$Sr measurements on selected  examples showing the superstructure caused by the smaller  rare-earth elements.

\section*{Results}
\subsection*{Crystal Structure and Stoichiometry}
\begin{figure*}[t]
\centering
\includegraphics[width=2\columnwidth]{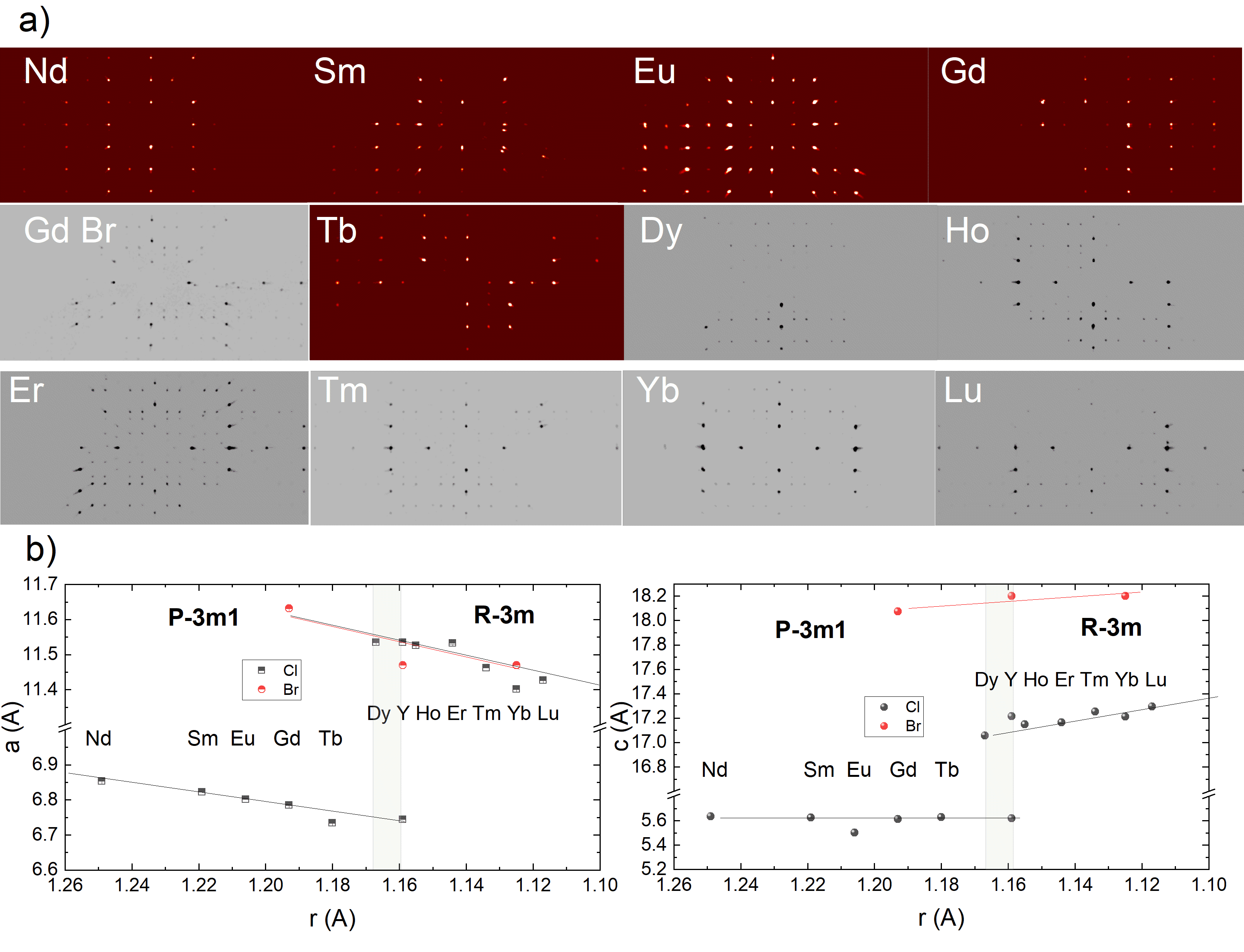}
\caption{\textbf{Single crystal diffraction. a)} Zonal diffraction maps of the (0 k l) sheet  of  a series of  $A$Cu$_3$(OH)$_{6+x}$(Cl,Br)$_{3-x}$ crystals.  For better visibility of the superlattice peaks of the triclinic crystals a black-white and for the subcell a red-white color view is chosen. In \textbf{b)} and \textbf{c)} the refined lattice parameters $a~=~b$ and $c$ are plotted as a function of  the radius of the $A-$site rare earth cations. Cl halogen atoms  are shown by black and Br atoms by red dots.
}
\label{XRD}
\end{figure*}

Figure \ref{XRD} displays zonal diffraction maps of
$A$Cu$_3$(OH)$_{6+x}$(Cl,Br)$_{3-x}$ crystals, where $A$ is a rare-earth cation.

$A$Cu$_3$(OH)$_{6+x}$(Cl,Br)$_{3-x}$ Kapellasite crystallizes for large $A-$site ions with a stoichiometry of $x=0$ as $A$Cu$_3$(OH)$_{6}$(Cl,Br)$_{3}$ in the trigonal space group $P\overline{3}m$1 (no. 164). The lattice constants range around $a~=~b\approx6.8$~\r{A} and $c\approx5.6$~\r{A} (see Fig. \ref{struc} a,c). For smaller $A$ site cations instead a $x=1/3$ variant crystallizes with the composition $A_3$Cu$_9$(OH)$_{19}$(Cl,Br)$_9$ with the trigonal space group $R\overline{3}$ (no. 148). Here, the lattice constants range around $a=b\approx11.5$~\r{A}, $c\approx17.3$~\r{A} as shown in Figure \ref{struc} b,d). The $x=1/3$ supercell is a structural subunit of the $x=0$ compound $A$Cu$_3$(OH)$_6$Cl$_3$, occurring from the substitution of 1/9 of Cl$^{-}$ by (OH)$^{-}$. Notably,  blends of mixed $x$ have been reported for Y, Cl \cite{Sun2017} and Y, Br \cite{Chen2020,Zeng2022,Liu2022,Lu2022}, but comparing the physical properties of the Cl phase \cite{Sun2017} with twofold magnetic entropy release at 15~K and 2.5~K to the end-members with single entropy release at either 15~K or 2.5~K \cite{Chatterjee2023}.  This findings rather indicates a inhomogeneous disordered phase mix whereas our optical transparent crystals always show clear  superstructure Bragg reflections.

\begin{figure}[h]
\centering
\includegraphics[width=1\columnwidth]{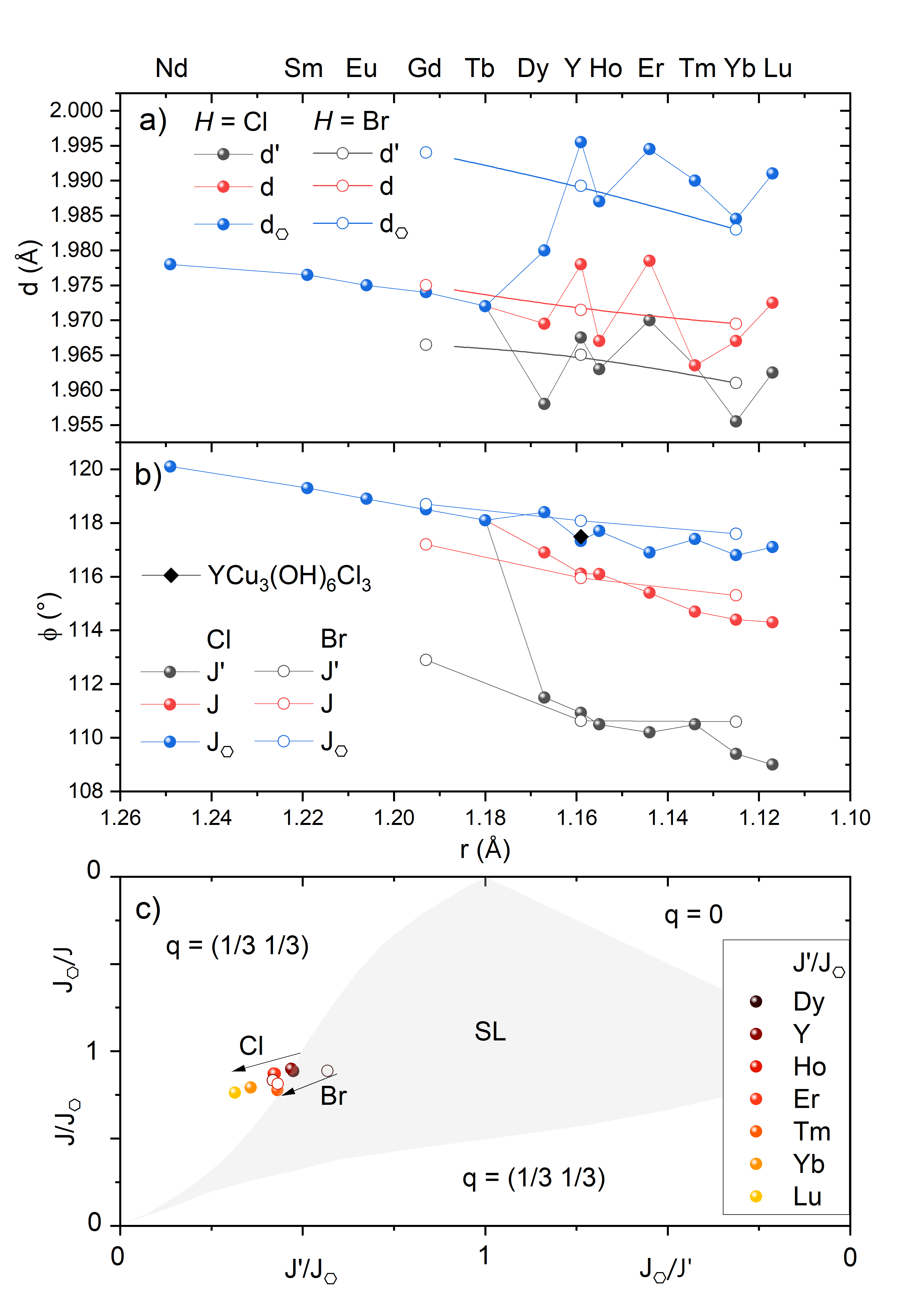}
\caption{\textbf{Superexchange. a)} Distance of the Cu~-~O bonds and \textbf{b)} Angle $\phi$ of the Cu~-~O~-~Cu bonds. As discussed in the text, the Cu1~-~O~-~Cu1 bond is denoted by $J_h$, the Cu1~-~O~-~Cu2 bond denoted by $J$ and the shortest bonded Cu2~-~O~-~Cu1 labeled by $J'$. \textbf{c)} Phase diagram of the distorted kagome for varying exchange interaction ratios. The exchange interactions $J_{h}$, $J$ and $J'$ and their ratios have been calculated from the bonding angles as discussed in detail in the text.
}
\label{angle}
\end{figure}

In Figure~\ref{XRD} a) we display (0 k l) zonal diffraction maps of Bragg reflections in red-white color of the subcell shown in Figure \ref{struc} a,c), which clearly differ to the supercell in black-white color, where clear additional Bragg reflections are observed. The $R\overline3$ superstructure realizes {partially} shorter bonds between the planes on substituting Cl$^{-}$ by (OH)$^{-}$ and hence the $A$ atoms start to move out of the kagome planes. This leads to a buckling of the kagome layers leaving behind two different Cu positions.  The buckling shortens one of the Cu~-~Cu bonds to  3.26 \AA~ (for Y), while the others remain at around 3.38 \AA. This results in Cu~-~O~-~Cu bonding angles of 110\degree and  116\degree and 117\degree.  The superexchange via Cu~-~O bonds is known to be strongly dependent on the bonding angle and is maximal for 180\degree. A phase diagram considering the change of exchange interactions has been theoretically investigated in Ref. \cite{Hering2022}. The phase diagram for isolated and hence 2D kagome planes has three scenarios: $q=0$, $q=(1/3, 1/3)$ order at the corners and a classical spin liquid in the center, which is sketched in Fig. \ref{angle} c). Notably, in the presently investigated systems the $A-$site ions consist of magnetic rare-earth ions. These, however, are far separated  and do not establish clear LRO Similar e. g. to the $AT$O$_3$ perovskites \cite{Catalano2018,Yen2007,Miyasaka2003}.  There with $A=$rare-earth and $T=$Ni, Mn, V the rare-earth ions stay paramagnetic and mainly control the magnetism of the transition metal $T$ sublattice, a similar scenario  proposed and discussed for the present compounds.


In Figure \ref{XRD} b) and Table \ref{latticeconst} we display and list the lattice parameters for varying $A-$site cations. With the reduction of the   atomic radius of the $A$-site cations $r$,  a compression of the in-plane lattice parameter $a$ takes place. Interestingly, the $c-$ lattice parameter stays nearly constant for $A$Cu$_3$(OH)$_6$Cl$_3$ until the $x=1/3$-superstructure with $A_3$Cu$_9$(OH)$_{19}$(Cl,Br)$_{8}$ is formed. Starting from $A$~=~Dy the $c$ lattice parameter increases with decreasing radius of the $A$ cation. We attribute this effect to the built-up of chemical strain induced by in-plane compression due to  the $A-$ site reduction rather than to isotropic chemical pressure. 
Notably, we already observed the structural transition for $A$~=~Dy, different to results reported earlier derived from investigations on a solid state sintered powder sample \cite{Fu2021}. On the other hand, as reported before for the Y system both structural variants may co-exist, dependent on the presence of water molecules \cite{Barthelemy2019}.  Fu \textit{et al.} utilized a water-free solid-state sintering technique of DyCl$_3\cdot6$H$_2$O 1:3 CuO, which can result in both structural variants, whereas  our  hydrothermal synthesis route clearly substitutes the $P\overline{3}m$1 phase. Nonetheless as discussed in the magnetism section below we find tiny impurities of the $P\overline{3}m$1 phase. We crosschecked the existence of the superstructure by combined Energy-dispersive X-ray spectra (EDX) and single crystal x-ray diffraction (XRD) analyses.

\begin{figure}[h]
\centering
\includegraphics[width=1\columnwidth]{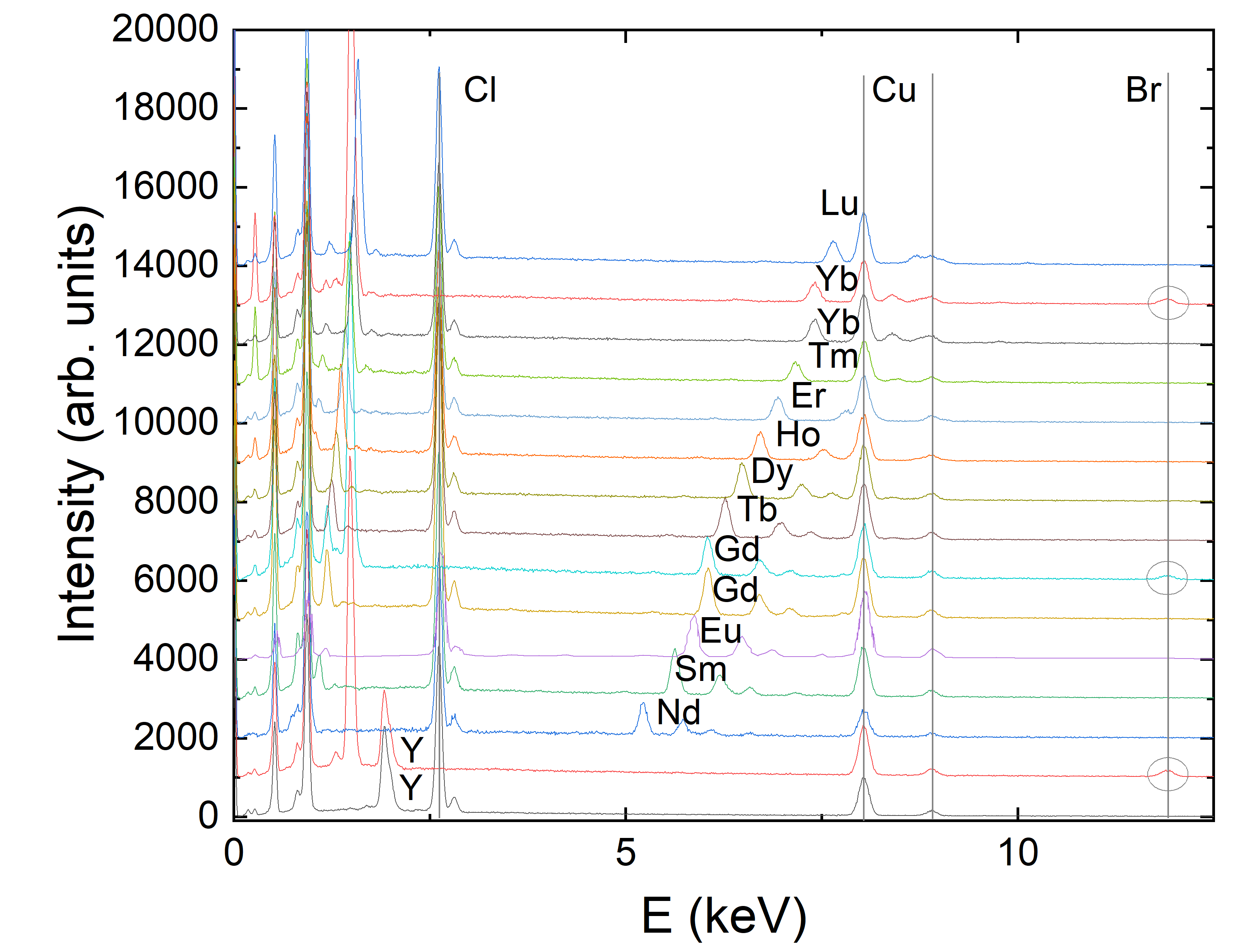}
\caption{\textbf{Selected EDS spectra of $A$Cu$_3$(OH)$_{6+x}$(Cl,Br)$_{3-x}$.} The data are presented in a waterfall diagram for the series with the $A-$site sorted by their number in the periodic table. For the halogen anion Br, we display the spectra always as a second spectra of the same $A-$site with the Br-line highlighted by a gray circle.
}
\label{EDX}
\end{figure}

We ensure phase pure high quality crystals of all rare-earth cations by EDX analysis. Examples of spectra are shown in Fig. \ref{EDX}, revealing highly stoichiometric crystals all within standard error bars of the expected formula, with a continuous shift of the L$_\alpha$ line.

We further determined the change of the atom positions with $A-$site variation as summarized in Tab. \ref{atompos}. With decreasing $A-$site rare-earth atom radius the atoms slowly move out of the kagome plane, in line with the sudden jump to a new cell appearing at around  $r(A)$~=~1.16 \AA\ when they have completely moved out of the  kagome layers.  The clearest trend is seen for the $z$-component of the atom positions which changes by about $\Delta\simeq0.4 \AA$, whereas the atom positions within the plane remain the same within experimental error. For the larger cell we see a continuation of this trend, that due to the increased amount of positions we do not list.

\begin{table}
\caption{\textbf{Lattice parameters.} For various $A=$ rare-earth ions as refined from single crystal XRD data with the zonal diffraction maps of the (h 0 l) sheet displayed in Fig. \ref{XRD}.}
\begin{tabular}{l|ll|lll}
 &   Cl &  & Br & \tabularnewline
 &   a (\AA) & c (\AA) & a (\AA) & c (\AA)\tabularnewline
\hline 
Nd & 6.8535(5) & 5.6352(4) &  & \tabularnewline
Sm & 6.8227(5) & 5.6249(5) &  & \tabularnewline
Eu  & 6.8014(9) & 5.5007(10) &  & \tabularnewline
Gd  & 6.7851(8) & 5.6118(9) &  11.6314(10) & 18.0728(19) \tabularnewline
Tb  & 6.7343(11) & 5.6287(9) & & \tabularnewline
Dy  & 11.5344(9) & 17.0578(13) &  & \tabularnewline
Y  & 6.7490(9) & 5.6244(11) &  & \tabularnewline
Y  & 11.5575(11) & 17.211(2) & 11.5945(5) & 18.2302(11)\tabularnewline
Ho  & 11.5259(8) & 17.1478(14) &  & \tabularnewline
Er & 11.533(4) & 17.164(4) &  & \tabularnewline
Tm  & 11.4618(10) & 17.252(2) &  & \tabularnewline
Yb  & 11.4011(9) & 17.2116(17) & 11.4703(6) & 18.2018(10) \tabularnewline
Lu  & 11.4270(7) & 17.2952(15) &  & \tabularnewline
\end{tabular}
\label{latticeconst}
\end{table}

\begin{table}
\caption{\textbf{Summary of the atomic positions for the $P\overline{3}m$1 structures}. The data were obtained from single crystal XRD refinements utilizing the zonal (h 0 l) maps  shown in Fig. \ref{XRD}. In the upper part the high symmetry positions are given. The lower table lists the free atom parameters for the compounds crystallizing in the space group $R\overline3$.}
\begin{tabular}{l|lllll}
 & x & y & z &  & \tabularnewline
\cline{1-4} 
$A$ & 1 & 1 & 0.5 &  & \tabularnewline
Cu & 0.5 & 0.5 & 0.5 &  & \tabularnewline
Cl & 1 & 1 & 0 &  & \tabularnewline
Cl & 2/3 & 1/3 & z$_{\rm Cl}$ &  & \tabularnewline
O & x$_{rm O}$ & y$_{\rm O}$ & z$_{\rm O}$ &  & \tabularnewline
 &  &  &  &  & \tabularnewline
 & Nd & Sm & Eu & Gd & Tb\tabularnewline
\hline 
z$_{\rm Cl}$ & 0.8647(4) & 0.8639(4) & 0.8651(15) & 0.8624(6) & 0.867(4)\tabularnewline
x$_{\rm O}$ & 0.6099(7) & 0.611(2) & 0.613(2) & 0.6154(11) & 0.620(3)\tabularnewline
y$_{\rm O}$ & 0.8049(3) & 0.806(4) & 0.8065(11) & 0.8077(5) & 0.8100(17)\tabularnewline
z$_{\rm O}$ & 0.6316(7) & 0.633(4) & 0.633(2) & 0.6332(11) & 0.625(5)\tabularnewline
\label{atompos}
\end{tabular}
\end{table}

As mentioned above, for rare earth cations with radii less than that of Y,  the $P\overline{3}m$1 crystal structure does not survive and it converts into a structure described by the space group $R\overline{\rm 3}$.  
However,  the kagome lattice is distorted, as already found for Y. Looking at the structural trends one may assume a continuous tuning of the magnetic ground state following the magnetic phase diagram of the distorted kagome layers (see Fig. \ref{angle}) \cite{Hering2022}.
The structural trends are best revealed by analyzing the Cu~-~O~-~Cu bonding angles. 
Starting from $A$~=~Dy,  we see a clear splitting of the Cu~-~O~-~Cu distances highlighted in Fig. \ref{angle} a):
In case of the $R\overline3$ superstructure we always detect three different Cu~-~O bonding distances: there is the Cu1 atom, which slightly moves away from (1/6, 1/3, 1/2) and Cu2 atom, which remains at a high symmetry position of (0 1/2 1/2). Cu1 creates a hexagon via the Cu1~-~O~-~Cu1 bond, with a large Cu~-~O distance extending from the initial Cu~-~O distance $d$ of 1.98~\r{A} to $d_h$ 1.99~\r{A} (see Fig. \ref{angle} a). These hexagons promote  the exchange path $J_h$ (orange line in Fig. \ref{struc} d) with the largest bond angle $\phi$ as plotted in  Fig. \ref{angle} b). Next is the Cu1~-~O~-~Cu2 bond  labeled as $d$, which is unchanged from the parent $P\overline{3}m$1 crystal structure showing a linear compressing trend and its Cu~-~O~-~Cu bond promotes the magnetic exchange path $J$ (blue line in Fig. \ref{struc} d). Finally, one bond shortens strongly labeled as $d'$. The Cu2~-~O~-~Cu1 angle is the relevant one, defining the degree of distortion. It has the lowest bond angle observed at around 110$\degree$, resulting in a decreased  exchange interaction $J'$ (green line in Fig. \ref{struc} d). In Fig. \ref{angle} c) we display the magnetic phase diagram of the distorted kagome lattice. Experimentally we showed that for Y we are very close to the SL border \cite{Chatterjee2023}. With disorder as evidenced by our powder samples \cite{Barthelemy2019}, or the mixed Y occupation found in the $A=$ Y and halogen = Br case with composition YCu$_3$(OH)$_{6}$Br$_{2}$[Br$_x$(OH)$_{1-x}$] \cite{Zeng2022,Zeng2024} the magnetic order is already destabilized and a SL state is realized. 

From our study of spin waves in the $A=$Y case using inelastic neutron scattering experiments, we found that $J_{\hexagon} \approx J = 140$ K and $J'=63\pm7$ K determined at a temperature of 1.55 K \cite{Chatterjee2023} (note the positive definition of AFM interaction). Using our lowest temperature diffraction result summarized in Fig. \cite{Chatterjee2023} we know the angles of $J_{\hexagon}$ and $J$ are 117 $\degree$ and 117.5 $\degree$ averaging to $117.25 \pm0.25 \degree$ and $J'$ has the angle $110.6\degree$. Consequently we can use a linear fit to establish the change of the exchange interactions for the angles that do vary within 108 to 120$\degree$ as observed in our XRD analysis.
We assume a simple linear dependence of the exchange interaction in relation to the Cu~-~O~-~Cu angle determining the strength of magnetic superexchange summarized in the equation
\begin{equation} 
J(K)\approx-1270+12\cdot\varphi[\rm{deg}]. 
\label{eq:AF}
\end{equation}
This results in a crossover from antiferromagnetic to ferromagnetic exchange at 105.83$\degree$ in agreement with findings e.g. for haydeiite \cite{Boldrin2015} stating a transition at around 105$\degree$.
For the closely related  kapellasite ZnCu$_3$(OH)$_6$Cl$_2$ with only one Cu~-~O~-~Cu bond of an angle of 105.5$\degree$ an exchange interaction of -12 K \cite{Kermarrec2014}  was found consistent with a value of -4.5 K derived from eq.(\ref{eq:AF}). Another example is the system herbertsmithite ZnCu$_3$(OH)$_6$Cl$_2$, where an exchange interaction of 180(10) K \cite{Khuntia2020} with a kagome bond angle of 119$\degree$ \cite{Kremer2024} was found comparing well with a value of 158 K obtained from eq.(\ref{eq:AF}).  
Applying the relationship to our diffraction results we find clear exchange interaction ratios that are displayed in the phase diagram in Fig. \ref{angle} c) extracted from Ref. \cite{Hering2022} indicating that chemical pressure moves  away from the border and thus the largest ion with the superstructure GdCu$_3$(OH)$_{6}$Br$_{3}$ could indeed reside in the SL region.

\subsection*{Magnetic Susceptibility}
To characterize the magnetic ground states we performed  magnetic susceptibility measurements. For the large $A^{3+}-$site ions Nd~-~Tb these measurements have been reported in Ref. \cite{Sun2017,Puphal2018,Fu2021}. As discussed in the Introduction, in case of the $P\overline{3}m$1 structure magnetic order from considerable DM-interaction occurs around 15~K, with a $q=0$ propagation vector resulting in a 120$\degree$ order as already known from the Ca and Cd variants. 

By decreasing the size of the $A^{3+}-$site cations,  we find our Dy-based single crystals have already formed the superstructure $R\overline{3}$ in contrast to other reports. The Dy-compound marks the first superstructure system with the largest $A-$ cation. This  magnetically completely non-characterized system will be discussed below. Next in size to Dy is Y. For this superstructure system various reports  on its magnetic properties exist \cite{Puphal2017,Barthelemy2019,Biesner2022,Chatterjee2023}.Around $T_N~=~2.5$ K a magnetic order transition is observed with only 10\% of the expected magnetic entropy \cite{Puphal2017}, and a small moment of only 0.38 $\mu_B$ was extracted from $\mu$SR \cite{Chatterjee2023}. In INS experiments  spin-waves at 2/3 and 4/3 were observed confirming the in-plane $q=(1/3 ~1/3)$ order in the kagome-layers as predicted by theory \cite{Hering2022}. Observation of a multi-center magnon further confirms a LRO scenario \cite{Biesner2022} supported by  \textsuperscript{1}H-NMR experiments which found a clear peak of the spin-lattice relaxation rate  $T^{-1}$  at $T_{\rm N}$ \cite{Wang2023}.  However, this order is not very pronounced in magnetization measurements, where only at lowest temperatures, far below the transition, several humps are seen. \cite{Biesner2022} Notably, in $M(H)$ data a plateau at around 0.2 $M_s$ is observed that is sharper in the $c-$direction, but similarly visible along $a$, as shown in \cite{Kremer2024,Biesner2022}.
The only other  superstructure $R\overline{3}$ system reported so far, is given by $A=$Gd  with Br as the halogen whereas the Cl system crystallizes with the $P\overline{3}m$1 structure.\cite{Cheng2022}  Here, no clear magnetic transitions are seen in the susceptibility. A peak in the specific heat is observed at around 2.5~K comprising an entropy of only 10~J/mol K$^{-1}$ corresponding to exactly 10\% of the maximum magnetic entropy of 103.73 J /mol$K^{-1}$, as found in the Y system.

\begin{figure}[h]
\centering
\includegraphics[width=1\columnwidth]{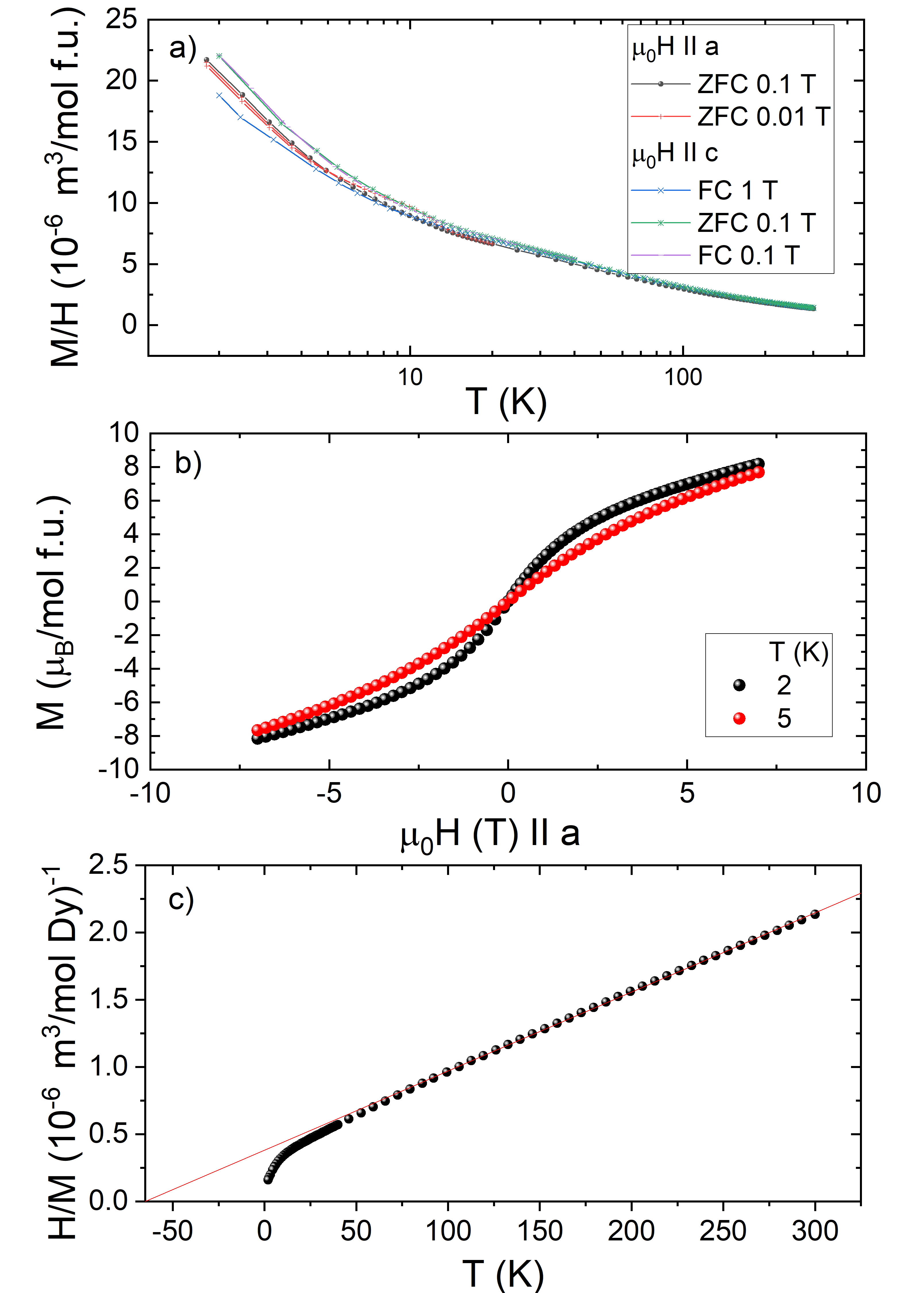}
\caption{\textbf{Magnetization and magnetic susceptibility data of Dy$_3$Cu$_9$(OH)$_{19}$Cl$_{8}$. a)} A semilog plot of the magnetization versus temperature in applied fields of 0.01-1~T both along the $a-$ and $c-$direction of Dy$_3$Cu$_9$(OH)$_{19}$Cl$_{8}$. \textbf{b)} Magnetization versus field at 2~K and 5~K for fields along the $a-$direction. \textbf{c)} Inverse of the magnetization versus temperature with a Curie-Weiss fit.
}
\label{DyM}
\end{figure}
Now we focus on the system Dy$_3$Cu$_9$(OH)$_{19}$Cl$_{8}$. In Figure \ref{DyM} a) we show the magnetic susceptibility versus temperature in a semi-logarithmic scale  for varying field strength applied both along $a-$ and $c-$ directions. Even with the semi-logarithmic representation the field dependence is small and neither magnetic transitions nor anisotropy is apparent. The magnetization versus field is plotted in Fig. \ref{DyM} b). At 7~T the magnetization starts to slightly level off around 8 $\mu_B$ per formula unit (f.u.), which is still far from the maximum moment of $M_s=gJ\mu_{\rm B}=(3\cdot10+9\cdot1)\mu_{\rm B}=39 \mu_{\rm B}$ that may be expected per formula unit for Dy$_3$Cu$_9$(OH)$_{19}$Cl$_{8}$.
The  observation of a subtle leveling off of the magnetization suggests a possible 1/3 plateau or 0.33 $M_s$ as observed for the Y-variants cases \cite{Kremer2024,Suetsugu2024}. 

The magnetic susceptibilities of rare-earth ions (except Gd) are largely determined by crystal electric field effects and Curie-Weiss behavior can at best be expected at very high temperatures \cite{Mugiraneza2022}. The fitted effective moments obtained from the high-temperature susceptibilities are commonly in fair agreement with the expected $g_J\sqrt{J(J+1)}$, where $g_J$ is the Land\'{e} $g$-factor and $J$ the total angular momentum. Derived Curie-Weiss temperatures from fits of the high-temperature susceptibilities are largely determined by crystal field splitting of the $J$ multiplets and consequently  do not allow meaningful conclusions about exchange interactions.

We show in Fig. \ref{DyM} c) the inverse of the susceptibility and apply a Curie-Weiss fit in the range of 100 - 300 K, yielding a subtly reduced Curie-Weiss temperature of -65(2) K compared to -100(4) K for Y \cite{Puphal2017}. The effective moment $\mu_{\rm eff}$ obtained from the slope of 0.0056(2) for the inverse susceptibility per Dy  amounts to $\sim$10.7(3)~$\mu_{\rm B}$,
in line with the expected value dominated by the rare earth moment as one Dy$^{3+}$ has effective moments of $\sim$10.6~$\mu_{\rm B}$ and 3 Cu$^{2+}$ cations each would contribute with $\sim$1.8~$\mu_{\rm B}$ leading to an expected effective moment of $\sqrt{10.6^2+3\cdot1.8^2}~\mu_B\approx11.05~\mu_B$ consistent with our findings.

\begin{figure}[h]
\centering
\includegraphics[width=1\columnwidth]{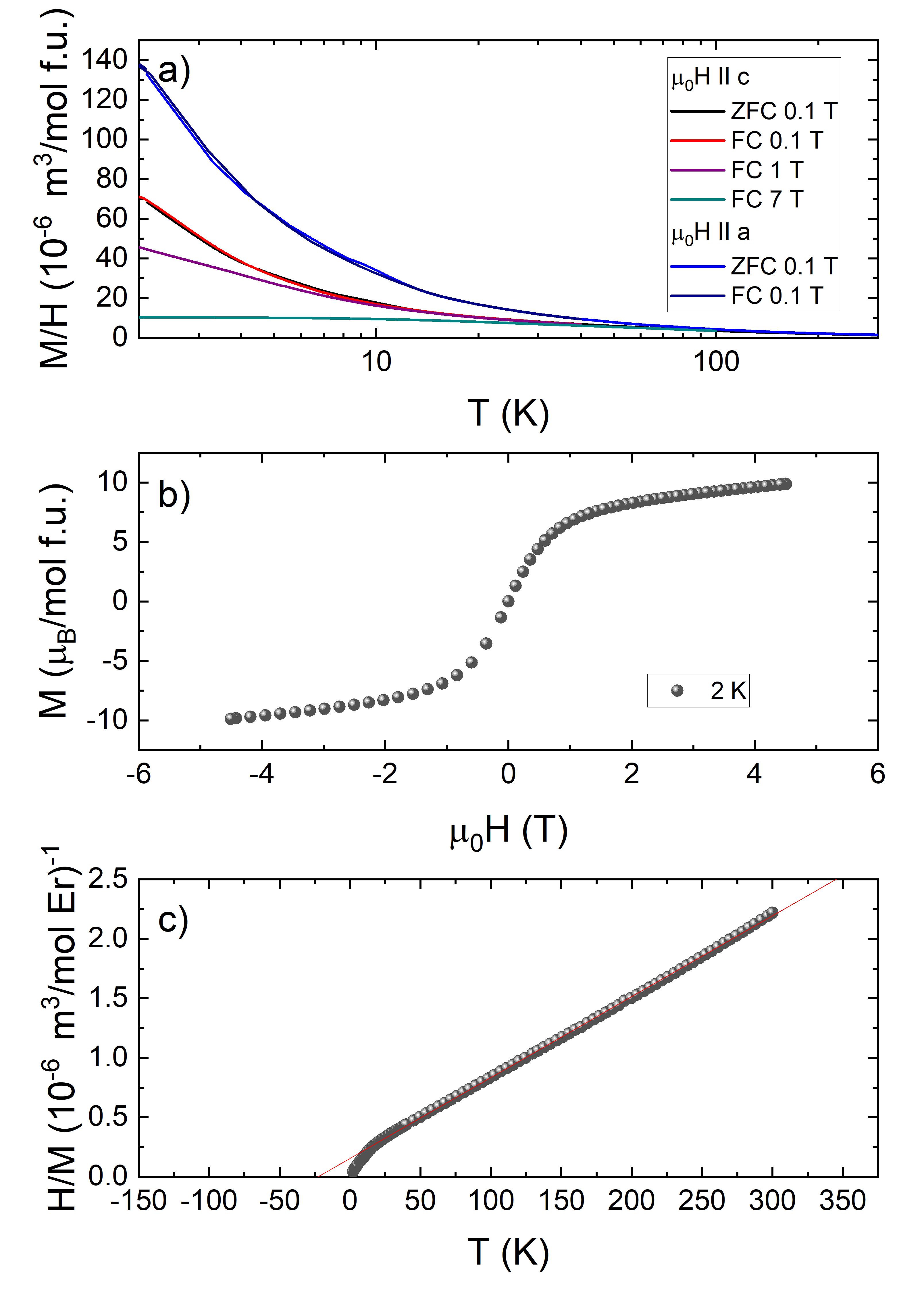}
\caption{\textbf{Magnetic data of Er$_3$Cu$_9$(OH)$_{19}$Cl$_{8}$  a)} Semilog plot of the magnetization versus temperature in fields of 0.01-1 T both along the $a-$ and $c-$direction of Er$_3$Cu$_9$(OH)$_{19}$Cl$_{8}$. \textbf{b)} Magnetization versus field at 2~K and 5~K for fields along the $a-$direction. \textbf{c)} Inverse of $M/H(T)$ versus temperature with a Curie-Weiss fit (solid red line).
}
\label{ErM}
\end{figure}

Next, in Figure \ref{ErM} a) we show the magnetic susceptibility versus temperature in a semi-logarithmic scale, for Er$_3$Cu$_9$(OH)$_{19}$Cl$_{8}$ at varying field strength applied both along $a-$ and $c-$ direction. Again we see no clear magnetic transition, but the field dependence and anisotropy is enhanced. The magnetization versus field is plotted in Fig. \ref{DyM} b), where each Er$^{3+}$ cation contributes with 9~$\mu_{\rm B}$. Similarly, around 7 T the magnetization levels off and we are close to a plateau, which would be more apparent at lower temperatures. The moment lies around 10 $\mu_B$, {comparing to} $M_s=gJ\mu_{\rm B}=(3\cdot9+9\cdot1)\mu_{\rm B}=36\mu_{\rm B}$, as expected for full polarization of the Er and Cu magnetic moments. Here, extrapolating the temperature dependence we can expect 12 $\mu_{\rm B}$ around 0.5 K stabilizing a 1/3-plateau or 0.33 $M_S$.

As shown in Fig. \ref{DyM} c) a Curie-Weiss law was fitted in the range of 100 to 300 K with a $\theta_W=-23(2)~$K. The slope of 0.0068(2) corresponds to an effective moment of $\sim$9.7(2)~$\mu_{\rm B}$ close to the expected value of $\sqrt{9.6^2+3\cdot1.8^2}~\mu_B\approx10.09~ \mu_B$ per Erbium atom.
\begin{figure}[h]
\centering
\includegraphics[width=1\columnwidth]{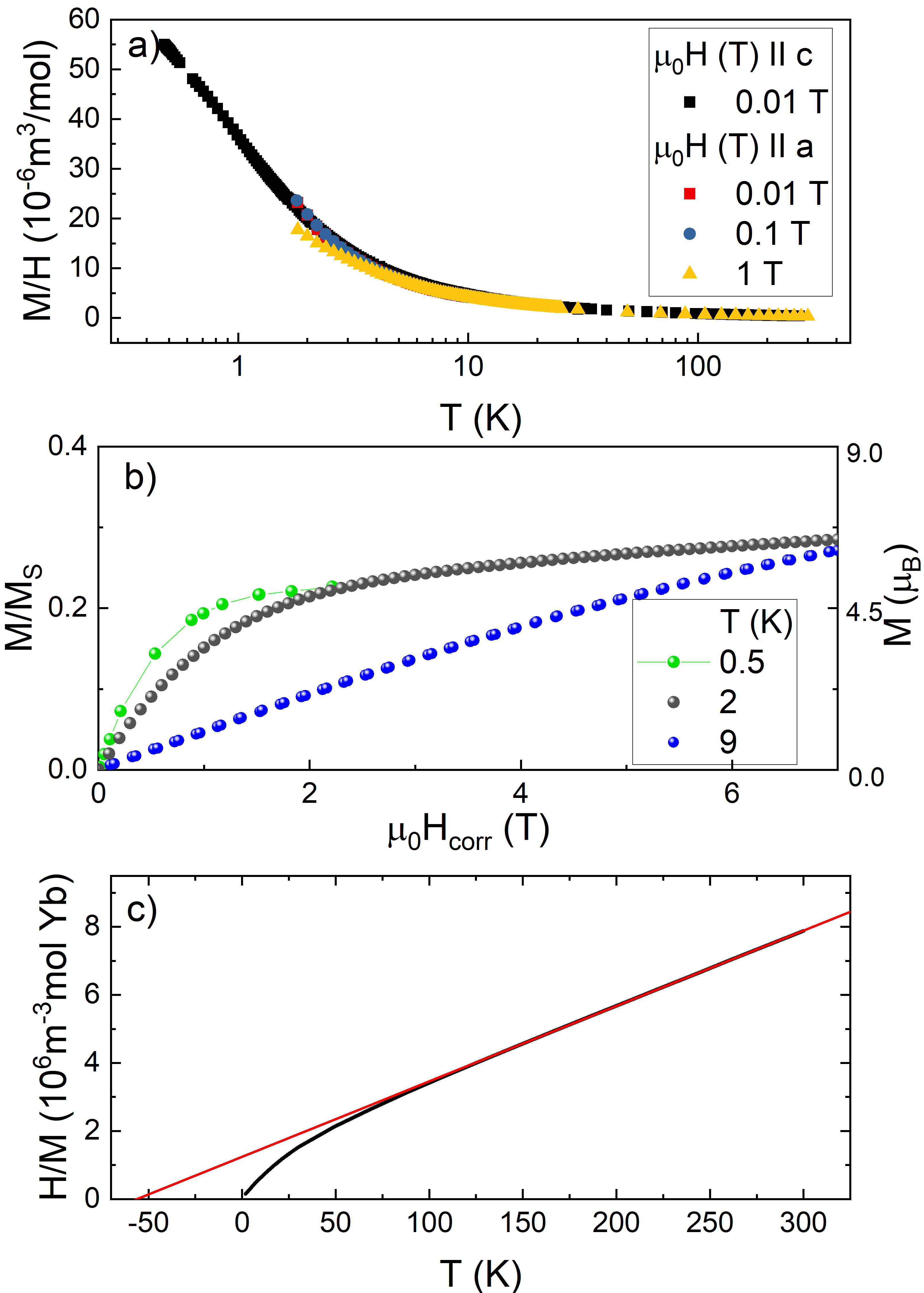}
\caption{\textbf{Yb$_3$Cu$_9$(OH)$_{19}$Br$_{8}$ magnetization. a)} Semilog plot of the Magnetization versus temperature in applied fields of 0.01-1 T  of Yb$_3$Cu$_9$(OH)$_{19}$Br$_{8}$ extending down to 400 mK. \textbf{b)} Magnetization versus field at 0.5~K, 2~K and 9~K for fields along the $a-$direction. \textbf{c)} Inverse of the magnetization versus temperature with a Curie-Weiss fit.
}
\label{YbM}
\end{figure}

Yb$_3$Cu$_9$(OH)$_{19}$Br$_{8}$ presents an interesting candidate as structurally it is very close to the Dirac spin liquid candidate YCu$_3$(OH)$_{6+x}$Br$_{3-x}$ \cite{Kremer2024,Jeon2024,Zeng2024}, and thus resides at nearly the same point in the phase diagram (see Fig. \ref{angle} c). Notably, with a similarly $S=1/2$ magnetic crystal field ground state of the rare-earth ion on the $A-$site it is worth to study its influence on the ground state properties.
In Figure \ref{YbM} a) we show the magnetic susceptibility versus temperature in a semi-logarithmic scale, for Yb$_3$Cu$_9$(OH)$_{19}$Br$_{8}$ at varying field strength applied both along $a-$ and $c-$ direction. Again we see no clear magnetic transition. The magnetization versus field is plotted in Fig. \ref{YbM} b).  The measurement at 0.5~K and 2~K show a clear plateau. The magnetic moment of the plateau lies around 6.5~$\mu_{\rm B}$ compared to $M_s=gJ\mu_{\rm B}=(3\cdot4+9\cdot1)\mu_{\rm B}=21\mu_{\rm B}$, as expected for full polarization of three Yb$^{3+}$ and nine Cu$^{2+}$ magnetic moments. Hence the moment of 6.5~$\mu_{\rm B}$ again best corresponds to a 1/3-plateau or 0.33$M_S$. We show a Curie-Weiss law fit in the range of 100 to 300 K in Fig. \ref{DyM} c) yielding $\theta_W=-56(1)$ K.
The effective moment amounts to ~$5.4(2)\mu_{\rm B}$ as compared to the expected $\sqrt{4.53^2+3\cdot1.8^2}\approx5.5~\mu_{\rm B}$ for the combination of one Yb$^{3+}$ and three Cu$^{2+}$ moments, well in line within experimental error.

Accordingly, all superstructure variants  around room exhibit a  temperature dependence  expected for a paramagnetic contribution from all magnetic moments. At low temperatures no clear magnetic transitions are apparent and the magnetic anisotropy is minimal as expected for the Cu  but not for the rare-earth cations hinting at their paramagnetic nature.  Magnetization plateaus at low temperatures are found, with Dy deviating due to disordered mixed crystals as discussed in the following.

\subsection*{Specific heat}
\begin{figure}[h]
\centering
\includegraphics[width=1\columnwidth]{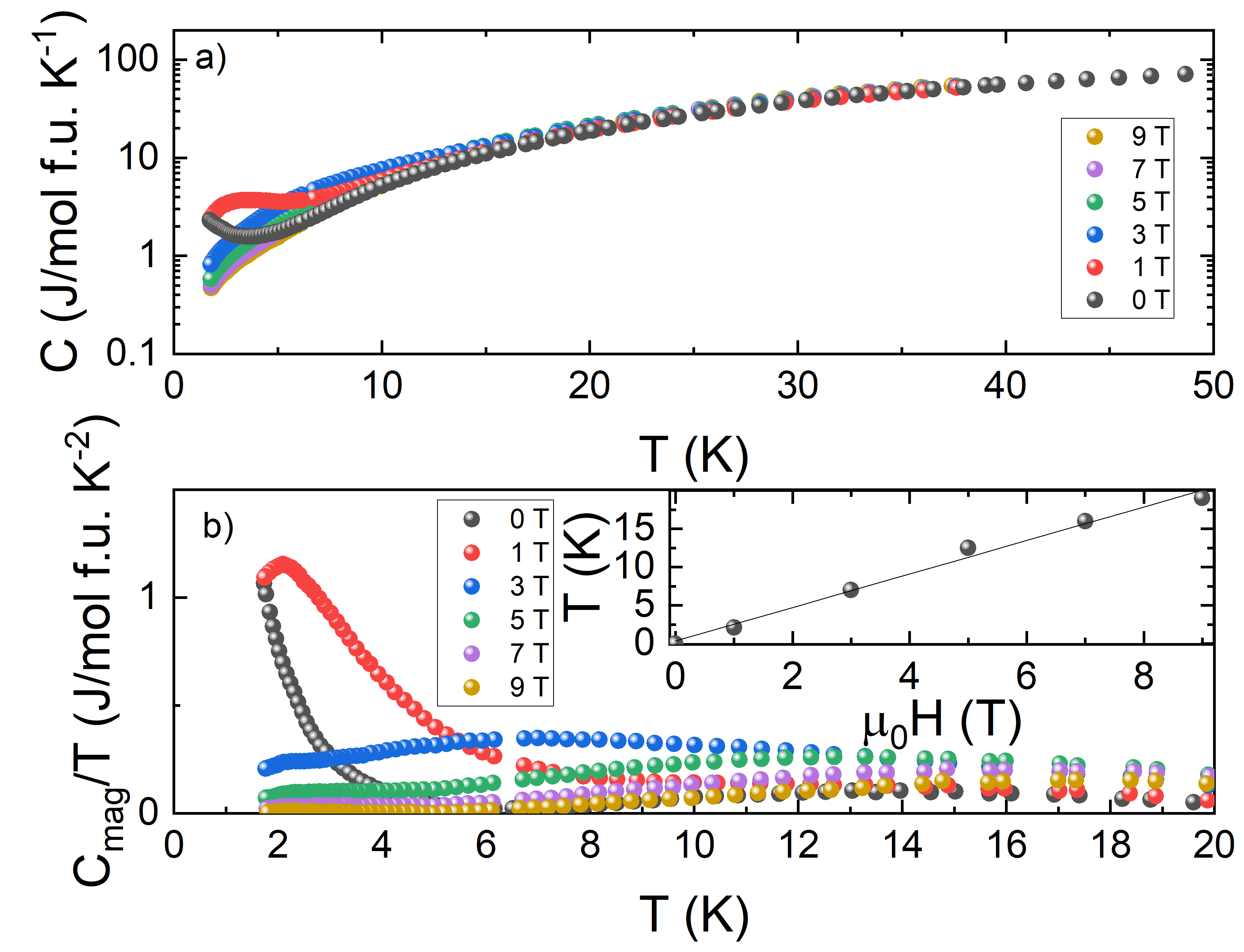}
\caption{\textbf{ Specific heat of Dy$_3$Cu$_9$(OH)$_{19}$Cl$_{8}$. a)} at 1.8 - 50 K in increasing magnetic field from 0~-~9 T of Dy$_3$Cu$_9$(OH)$_{19}$Cl$_{8}$ single crystals coaligned with field applied along the $c$-direction. \textbf{b)} Magnetic contribution of the specific heat, after subtraction of a phonon contribution  in the temperature range from 1.8 - 20 K. The inset shows the temperature of the maximum versus the applied field.
}
\label{Dy}
\end{figure}

Next, we study the specific heats starting with Dy$_3$Cu$_9$(OH)$_{19}$Cl$_{8}$. In Fig. \ref{Dy} a) the specific heat versus temperature of Dy$_3$Cu$_9$(OH)$_{19}$Cl$_{8}$ is displayed in a large temperature range  for varying magnetic fields with a semi-logarithmic scale to visualize the maximum developing at around 2.5~K. This maximum strongly shifts with the application of external fields and broadens.
In Fig.~\ref{Dy} b) we show the low temperature part of the magnetic contribution to the specific heat divided by temperature $C_{mag}/T$.  The magnetic contribution was obtained by subtracting the phonon contribution approximated assuming a Debye temperature of 603~K estimated by a linear fit of $C/T$ vs $T^2$ in the temperature range of 5 to 20 K. Early reports of results reported for powder samples of Dy$_3$Cu$_9$(OH)$_{19}$Cl$_{8}$ showed it to crystallize in the $P\overline{3}m$1 structure known to exhibit LRO around 15~ K. While our crystals showed superstructure reflection, we still find considerable entropy release around 15~K, even in zero field, hinting at impurities of DyCu$_3$(OH)$_6$Cl$_3$ or at a crystal   with mixed $x$ values leading to site disorder on the Dy site. Notably, the release of magnetic entropy around 2.5 K is hard to estimate since the peak is not yet at its maximum even at our lowest temperature of 1.8~K. If we extrapolate the data in an external field of 1~T assuming a $T^2$ behavior fitting to spin-wave contributions found for the Y case, we find an entropy of around 4.5 J/mol K$^{-1}$. The magnetic entropy can be estimated via $\Delta S=R$ln$(2J+1)$, with 3 Dy$^{3+}$ and 9 Cu$^{2+}$ we have a total entropy of $\Delta S=3R$ln$2+9R$ln$2=69$ J/mol f.u. K$^{-1}$, assuming a crystal field doublet as ground state of the Dy$^{3+}$ cations. Hence only 6.5\% of entropy is released in this magnetic transition. This is less than oberved for the Y system which releases 10\% \cite{Puphal2017,Chatterjee2023}. However, for Dy additional entropy is already released around 15 K in a broad peak. While our single crystal XRD clearly show superstructure reflexes, the entropy release around 15~K is ascribed to a mixed system of $x=0$ $P\overline{3}m$1 and $x=1/3$ $R\overline{3}$ leading to a variation of Cl and OH content as is similarly observed for YCu$_3$(OH)$_{6+x}$Cl$_{3+x}$ \cite{Chatterjee2023}.

At higher fields a splitting of the Kramer's doublet ground state of Dy$^{3+}$ of the rare-earth e.g. Dy$^{3+}$ ion is expected and for field smaller than the crystal field splitting  a shift $\Delta T$ of the specific heat maximum linear with field may be expected. Via a straightforward linear fit of  $\Delta E=k_{\rm B}\Delta T\propto g\mu_{\rm B}$ one can estimate the $g$-factor of the crystal field ground state. This is shown on the inset of Fig. \ref{Dy}, extracted from the maximum in the magnetic contribution in $C_{mag}/T$ of the main panel.
{The slope corresponds to a $g$-factor of 3.25.}

\begin{figure}[h]
\centering
\includegraphics[width=1\columnwidth]{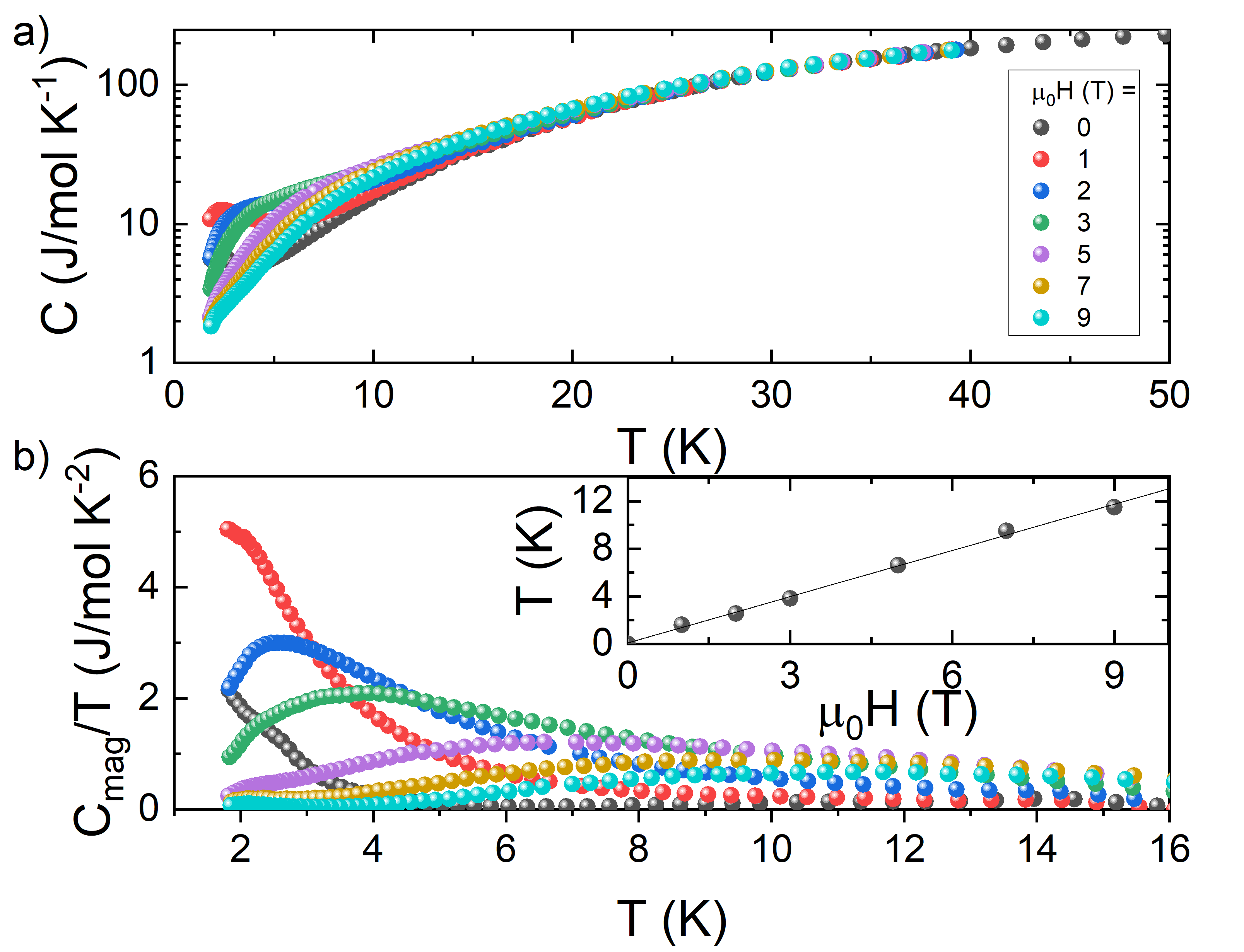}
\caption{\textbf{a) Specific heat of Er$_3$Cu$_9$(OH)$_{19}$Cl$_{8}$. a)} with field applied along the $c$-direction  at 1.8 - 50 K in increasing magnetic fields from 0~-~9~T. \textbf{b)} Magnetic contribution to the specific heat of Er$_3$Cu$_9$(OH)$_{19}$Cl$_{8}$, after subtracting a phonon subtraction, in the temperature range  1.8 - 16~K. The inset shows the temperature of the maximum versus the applied field.
}
\label{Er}
\end{figure}

The specific heat data for Er are summarized in Fig. \ref{Er} a) as a function of temperature in a large range for varying fields with a semi-logarithmic scale, to visualize the maximum developing around 2.5~K. As for Dy before the  maximum strongly shifts with the application of external fields and broadens, but notably less as in accordance with the reduced Land\'{e} $g$-factor of Er$^{3+}$.
In Fig. \ref{Er} b) we show the low temperature part of the magnetic contribution in the specific heat divided by temperature $C_{mag}/T$.  As before, the magnetic contribution was estimated by subtracting the phonon contribution with a Debye temperature of 628~K estimated by a linear fit of $C/T$ vs $T^2$ in the temperature range of 5 to 20~K. Notably the release of magnetic entropy is even harder to estimate since our peak  even at 1.8 K has not yet reached its maximum and even with 1~T external field we do not see a pronounced maximum. If we extrapolate the 2~T data assuming a $T^2$ behavior fitting to spin-wave order found for the Y case, we find an entropy of around 11.5 J/mol f.u. K$^{-1}$ corresponding to $\approx$17\% of $3R$ln$2+9R$ln$2=69$ J/mol f.u. K$^{-1}$ again assuming a Kramers doublet as crystal field ground state for Er$^{3+}$ and $S=1/2$ of Cu$^{2+}$. {The entropy release is thus similar to that of the Y based system with where a value of 10\% was observed.}

Displayed in the inset of Fig. \ref{Er} is the shift of the maximum of $C_{mag}/T$ with external magnetic field.
From a linear fit of the shift of the maximum we estimate the $g$-factor \$of the Kramers doublet ground state to 1.93.

\begin{figure}[h]
\centering
\includegraphics[width=1\columnwidth]{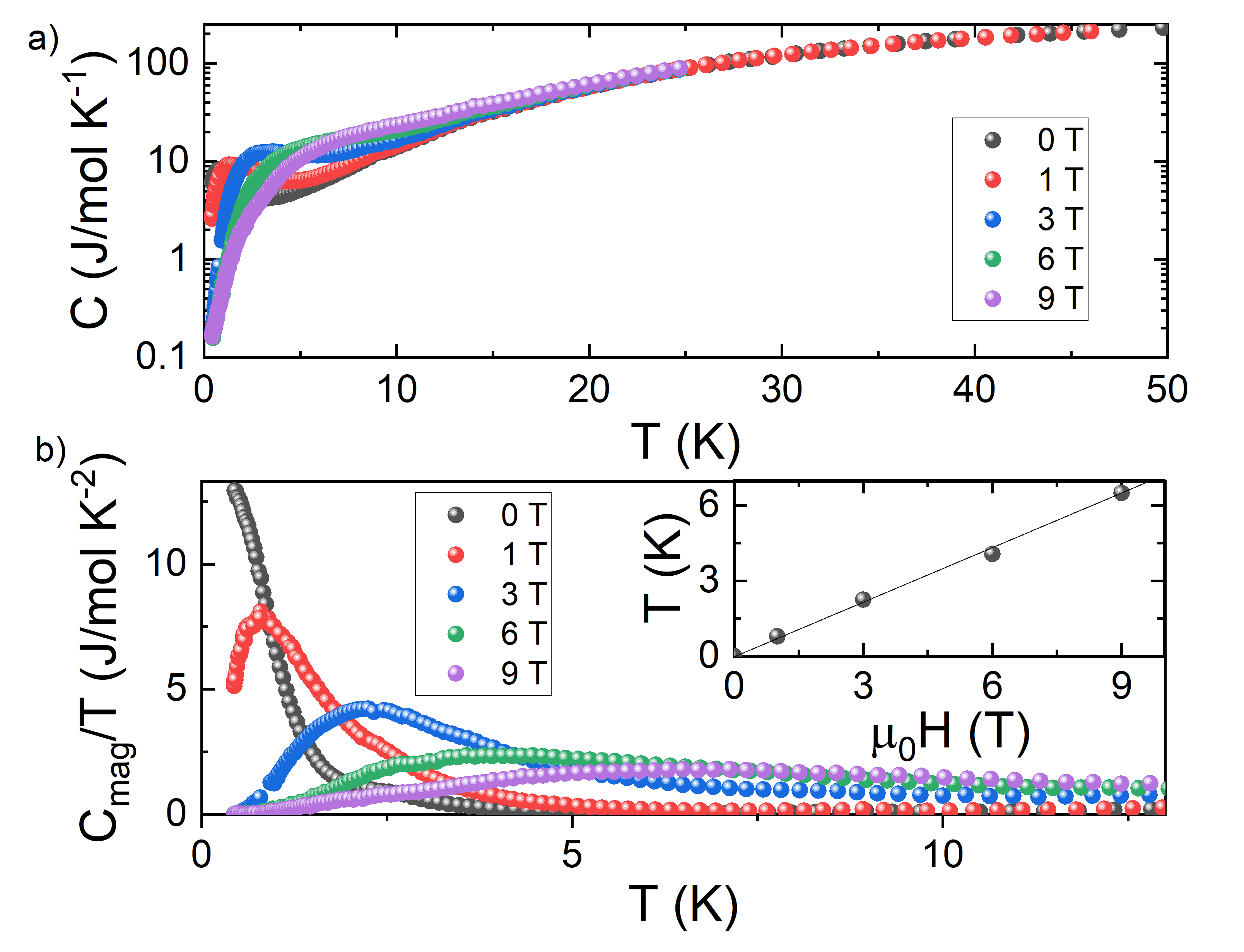}
\caption{\textbf{Specific heat of Yb$_3$Cu$_9$(OH)$_{19}$Br$_{8}$ a) }  at 0.4 - 50 K in increasing field from 0-9~T of Yb$_3$Cu$_9$(OH)$_{19}$Br$_{8}$ single crystals coaligned with field applied along the $c$-direction. \textbf{b)} Magnetic contribution of the specific heat obtained after subtracting a phonon contribution. The inset shows the maximum in temperatures versus the applied magnetic field.
}
\label{Yb}
\end{figure}

The specific heat versus temperature for Yb is displayed in Fig. \ref{Yb} a). The specific heat has been measured from 50~K down to 0.4~K and plotted for varying fields on a semi-logarithmic scale to visualize the maximum, again developing at around 2.5~K. Still this maximum strongly shifts with the application of external fields and broadens, but notably even less as in accordance with the again reduced Land\'{e} $g$-factor of Yb$^{3+}$.
In Fig. \ref{Yb} b) we show the low temperature part of the magnetic contribution in the specific heat divided by temperature $C_{mag}/T$. $C_{mag}/T$ in the temperature range from 0.4~K to 20~K was obtained as before by subtracting a phonon contribution to the specific heat assuming a Debye temperature of 592 K from a linear fit of $C/T$ vs $T^2$ in the temperature range of 5 to 20 K.  We note that the nuclear contributions to the specific heat of Yb mainly starts below 0.4~K and hence its contribution can be neglected. The release of magnetic entropy is even with an extended temperature range hard to estimate since the peak is not yet at its maximum. But in an external field of 1~T a peak clearly develops. If we extrapolate the 1~T data assuming a $T^2$ behavior as found for Y assuming either spin-wave order, or also fitting to a Dirac SL state found for the Br case, we obtain an entropy of around 12.5 J/mol K$^-{1}$, with 10 J/mol K $^-{1}$ released down to 0.4~K. With Yb$^{3+}$ we expect an entropy of $3R$ln$2R$ln$2=69$ J/mol K $^-{1}$ amounting to 18\%, again amounting to a similarly low entropy release around 10\%.

Looking at the splitting of the Kramer's doublet ground state of the rare earth of Yb$^{3+}$ the shift of the specific heat maximum still follows the expected linear trend with field. Via a linear fit shown in the inset of Fig. \ref{Yb} we estimate the $g$-factor of the Kramers doublet crystal field ground state to be 1.05.

In summary for specific heat we see an uniform reduced release of the magnetic entropy at a transition around 2.5~K for the superstructure systems.

\subsection*{$\mu$SR Measurements}
We investigated the magnetic ground state of Yb$_3$Cu$_9$(OH)$_{19}$Br$_{8}$  in more detail by performing $\mu$SR measurements.
Above 2.5~K, the $\mu$SR relaxation hardly depends on temperature nor on the morphology of the sample. Powder or single crystal samples showed very similar behavior.\cite{Barthelemy2019,Chatterjee2023}. When the system is paramagnetic, we expect that we are in the fast fluctuation limit, and the relaxation is dominated by the quasi static, weak and random, nuclear fields. We follow a description of the magnetism as discussed in Ref. \cite{Barthelemy2019,Chatterjee2023}. Here, the muon decay asymmetry at high temperature is described by
 \begin{equation}\label{eq-2}
 a_0 P_{para}(t)=a_0 \left[ f P_{OH}(t) + (1-f)KT_{\Delta_{Br}}(t) \right]\\
 \end{equation}
 where
\begin{multline}
\label{eq-3}
P_{OH}(t) = e^{-\frac{ \Delta^2_{OH}t^2} { 2}} {\bigg[}\frac{1}{6}+\frac{1}{3}\cos{{\bigg(}\frac{\omega_{OH}t}{2}{\bigg)}}+\\
\frac{1}{3}\cos{{\bigg(}\frac{3\hspace{1mm}\omega_{OH}t}{2}{\bigg)}}+\frac{1}{6}\cos{{\bigg(}{\omega_{OH}t}{\bigg)}}{\bigg]}.\
\end{multline}
$P_{OH}(t)$ characterizes the  formation of \(\mu\)OH complexes with a pulsation $\omega_{OH} = \frac{\hbar \mu_0 \gamma_\mu\gamma_H}{4\pi d^3}$ which depends on the $\mu$-H distance $d$, the gyromagnetic ratios $\gamma_H=267.513 $ Mrad/s/T and $\gamma_\mu=851.616 $ Mrad/s/T respectively for protons and muons, and a Gaussian broadening $\Delta_{OH}$ due to the other surrounding nuclear spins. The second term in Eq.~\ref{eq-2} is a static Kubo-Toyabe relaxation standing for the minority fraction $1-f$ of muon stopping sites, likely close to the Br$^{\textbf{-}}$ ions, where they experience a Gaussian distribution of static nuclear fields with a width $\Delta_{Br}$. The parameter $a_0$ is the initial muon decay asymmetry in our experimental conditions. 

\begin{figure}[h]
\centering
\includegraphics[width=1\columnwidth]{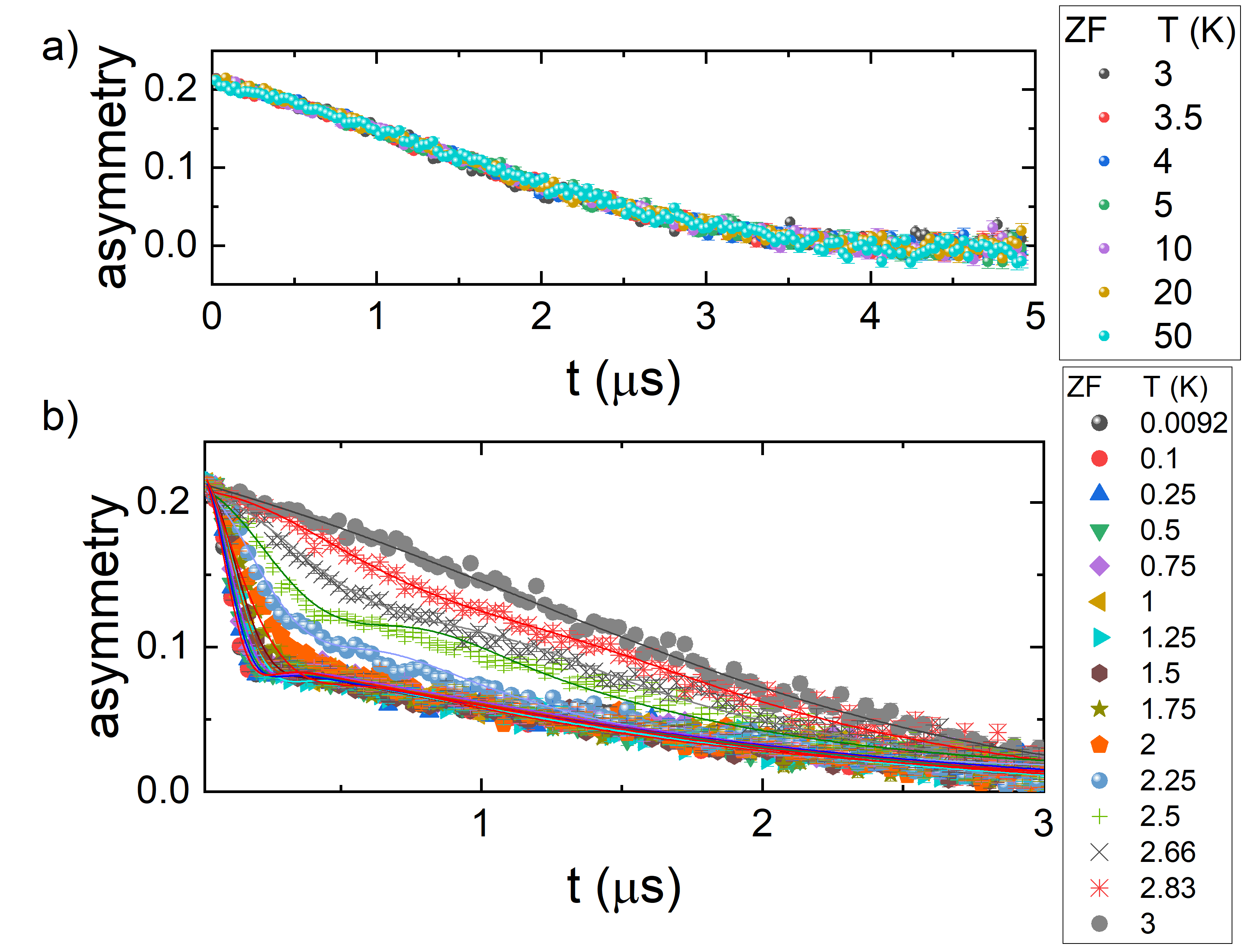}
\caption{\textbf{Zero field $\mu$SR} spectra of Yb$_3$Cu$_9$(OH)$_{19}$Br$_{8}$. a) Zero field high-temperature muon spin asymmetry in Yb$_3$Cu$_9$(OH)$_{19}$Br$_{8}$ for decreasing temperatures from 50~K to 3~K. \textbf{b)} Zero field muon spin asymmetry moving from the paramagnetic 3~K state to partial static order at low temperatures. The lines represent fits following the equation \ref{eq-4}
}
\label{zf}
\end{figure}

In Figure \ref{zf} a) we show the high temperature paramagnetic regime that is temperature independent consistent with a paramagnetic ground state in zero field for Yb$_3$Cu$_9$(OH)$_{19}$Br$_{8}$ down to 3~K. The nuclear relaxation parameters in Eq.~\ref{eq-2} and Eq.~\ref{eq-3} evaluated by fitting the ZF muon asymmetry at 3~K are presented in table~\ref{Tab:1} with values quite similar to related systems \cite{Barthelemy2019,Chatterjee2023}.  \\
\begin{table}[h]
\centering
     \begin{tabular}{|c|c|}
\hline
  Static parameters        & Yb$_3$Cu$_9$(OH)$_{19}$Br$_8$\\
\hline
    $f\%$   &  75.00$\pm$1.00\\
\hline
 $\omega_{OH}$~(Mrad.s$^{-1}$) &0.56$\pm$0.02\\
\hline
         d~(\AA) & 1.63$\pm$0.02\\
\hline
         $\Delta_{Br}~(\mu s^{-1})$ & 0.85$\pm$0.02\\
\hline
      $\Delta_{OH}$~($\mu$s$^{-1}$) & 0.21$\pm$0.02\\
\hline
      \end{tabular}
\caption{\textbf{Static nuclear parameters.} Derived from high temperature fit of the ZF asymmetry with Eq.~\ref{eq-2}.}
     \label{Tab:1}
   \end{table}

Below 3~K we see a sharp release in entropy, as shown in Fig. \ref{Yb}, and consequently a change in asymmetry attributed to a continuous increase of freezing spins visible in Fig. \ref{zf} b).
Hence upon cooling, the relaxation from the  electronic spins increases progressively and a strongly damped oscillation develops. Notably in case of Y$_3$Cu$_9$(OH)$_{19}$Cl$_8$ a dip was still visible, while here for Yb$_3$Cu$_9$(OH)$_{19}$Br$_8$ we cannot see even a single oscillation. Oscillations would indicate a magnetic transition in the single crystal samples. 
The ordered magnetic fraction can be described using
\begin{equation} \label{eq-4}
a_0\cdot P_f(t)=a_0\left[\frac{2}{3}\cos{(\omega_ft+\phi)}e^{{\frac{\sigma^2 t^2}{2}}}+\frac{e^{-\lambda_ft}}{3}\right]  
\end{equation}   
which accounts for a magnetically frozen spins with an average internal field at the muon sites $B_{int}=\omega_f/\gamma_\mu$.  The damping parameter $\sigma$ encodes the width of the distribution of these internal fields. The last exponential term accounts for residual fluctuations in the frozen phase. 

\begin{figure}[h]
\centering
\includegraphics[width=1\columnwidth]{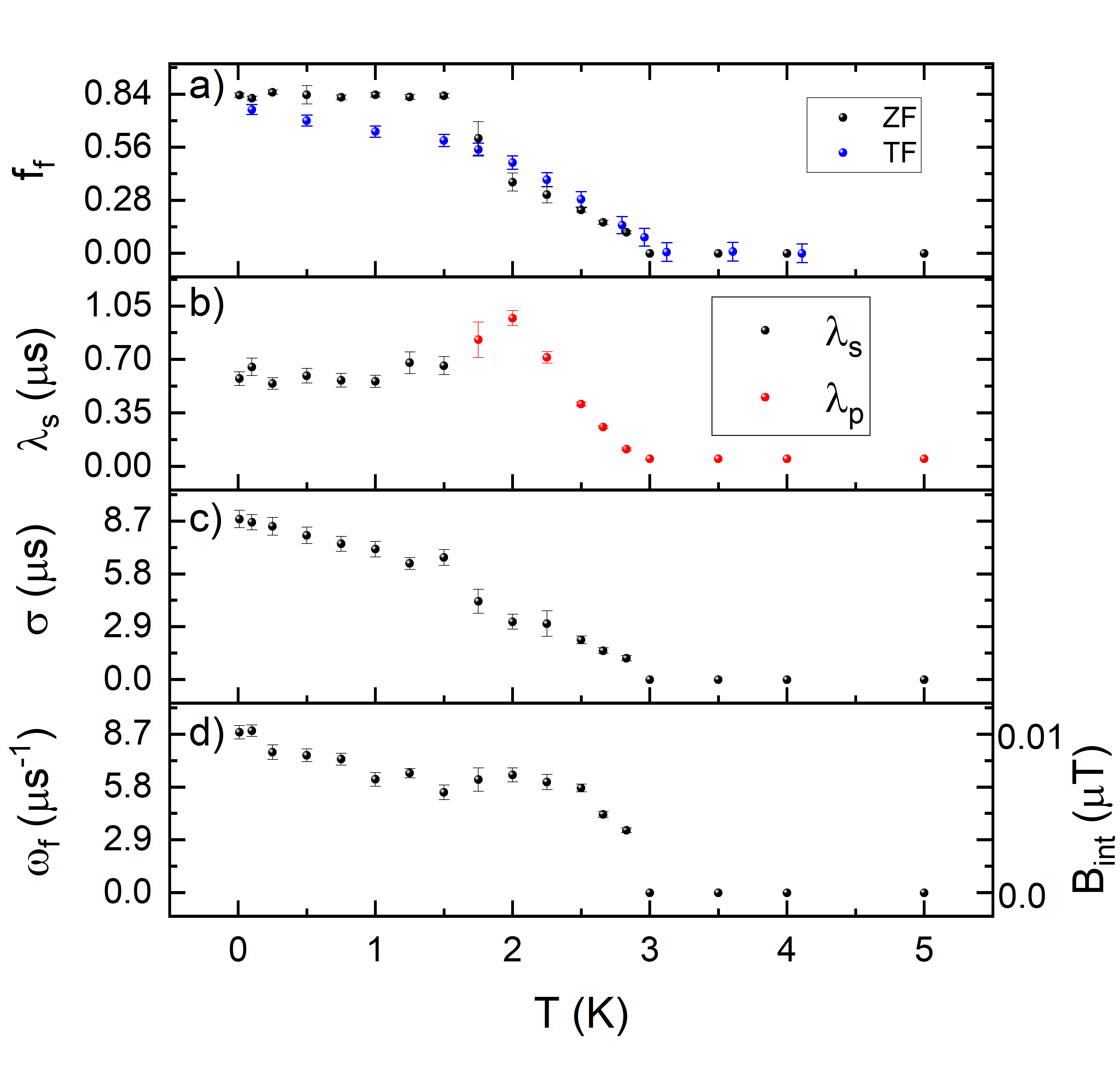}
\caption{\textbf{Results of the $\mu$SR analysis.} Fitting parameters of the ZF $\mu$SR data shown in Fig.\ref{zf} using equation \ref{eq-5} with fixed parameters from table \ref{Tab:1}. \textbf{a)} frozen fraction both extracted from ZF as well as the asymmetry change in a weak transverse field shown in Fig.\ref{wtlf}. \textbf{b)} the relaxation rates in the paramagnetic
 ($\lambda_p$) and frozen ($\lambda_f$) phases. \textbf{c)} the damping rate $\sigma$ and \textbf{d)} the frequency $\omega_f$ reflecting the internal field magnitude shown on the right axis.
}
\label{zffit}
\end{figure}

In order to fit the asymmetry over the whole temperature range, we combine both equations \ref{eq-2} and \ref{eq-4} in
\begin{equation} \label{eq-5}
a_0 P(t)=a_0 \left[ f_f P_f(t) + (1-f_f) P_{para}(t) e^{-\lambda_pt} \right] 
\end{equation}
with a switching parameter $f_f$ that tracks the frozen volume fraction from fully frozen for $f_f=1$ to fully paramagnetic for $f_f=0$. All the nuclear parameters are kept constant, so that the only varying parameters shown in Fig.~\ref{zffit} are the frozen fraction $f_f$, the oscillation frequency and its damping rate, and the relaxation rate $\lambda_p$ or $\lambda_f$ for the paramagnetic or the frozen phase. Using this equation we can describe the entire temperature range as shown in Fig. \ref{zf} b). Notably unlike for the Y compound \cite{Chatterjee2023} we find no fully ordered state of $f_f=1$. Similarly the transition is less pronounced in the relaxation rates. This is in accordance with the absence of a clear dip in the asymmetry and thus our data rather point to an inhomogeneously frozen state than to clear LRO for Yb$_3$Cu$_9$(OH)$_{19}$Br$_{8}$. 

\begin{figure}[h]
\centering
\includegraphics[width=1\columnwidth]{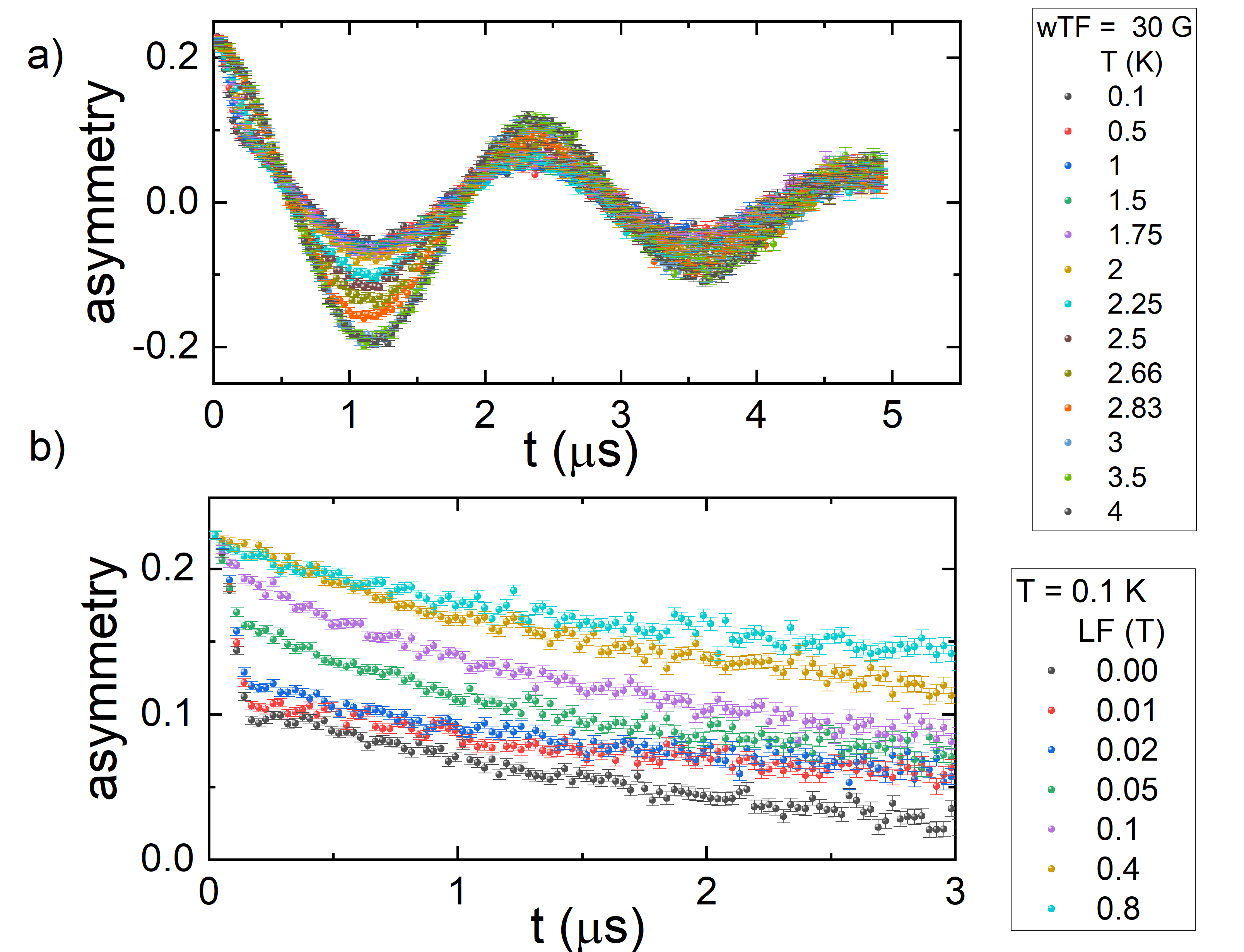}
\caption{\textbf{$\mu$SR in external field. a)} Muon spin asymmetry in a weak transverse field (wTF) of 30~G for Yb$_3$Cu$_9$(OH)$_{19}$Br$_{8}$ for decreasing temperatures from 4~K to 0.1~K. \textbf{b)} Longitudinal field (LF) muon spin asymmetry in the static low temperature region for the field range of 0 to 0.8~T.
}
\label{wtlf}
\end{figure}

Furthermore, we measured the sample with the application of a weak transverse field (TF), which causes the muon spins to precess in the field. The change of the amplitude, hence the asymmetry with temperature gives additional insights into the fraction of frozen spins. We fitted the polarization in a TF of 30~G (see Fig. \ref{wtlf} a) with a Gaussian damped cosine function representative for a Gaussian distribution of internal fields and show the resulting amplitude change i.e. the frozen fraction in Fig. \ref{zffit} a) as blue dots. Under this external field we find the same increase of frozen spins around the 3 K transition. However, in ZF we find an additional increase below 1.5 K that is absent in TF. Furthermore, we applied a longitudinal field (LF) at different temperatures. We show the change with increasing LF up to 0.8~T in Fig. \ref{wtlf} b) at 0.1~K. Due to the paramagnetic rare-earth ion we need enhanced external field to decouple our spins. Decoupling is reached at large external fields of 0.4 T, quite unlike the Y case. The remaining slight slope of the data at the highest fields indicates the presence of persistent dynamics even at lowest temperatures. Notably, there are  3 possible reasons for the persistent spin dynamics: 1) fluctuations in the non-frozen fraction 2) fluctuations of the Yb moments 3) fluctuations of the Cu system, i.e. SL like behavior. Here future neutron experiments will shed light on these complex systems, as so far no magnetic Bragg peaks were observed due to the low Cu moment, large rare-earth moments could enable a magnetic structure solution in case of LRO. Furthermore a series of spin wave studies could shed light on the chemical pressure control in these systems.

 To summarize, $\mu$SR measurements indicate a complex magnetic ground state. Already, for Y we find a strongly reduced moment, but clear signs of LRO \cite{Chatterjee2023}. Here, we are even closer to the SL state and hence a mix of frozen and fluctuating spins remain down to lowest temperatures likely only realizing short range order.

\subsection*{Phase Diagram}
\begin{figure}[h]
\centering
\includegraphics[width=1\columnwidth]{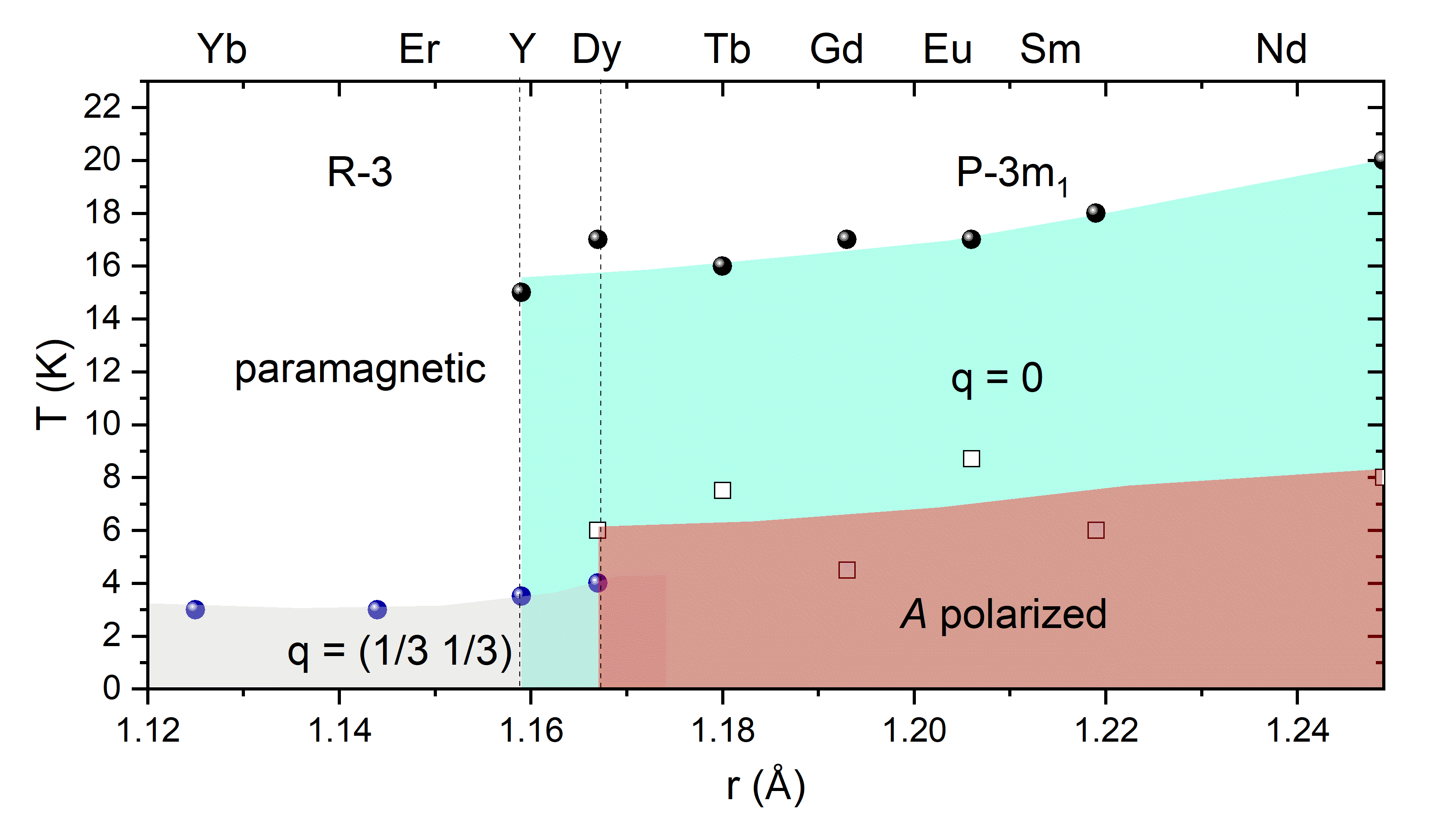}
\caption{\textbf{Magnetic phase diagram.} Magnetic ordering temperature versus ionic radius of the ion $A$ for the halide ion Cl. Given are the structures, in the white paramagnetic state and the ordering transitions towards q = 0 (turquoise) and in-plane q = (1/3 1/3). The red area highlights an additional transition, where the magnetic rare earth ion $A$ is polarized.
}
\label{dia}
\end{figure}

Finally, we summarize and construct the magnetic phase diagram depicted in Fig. \ref{dia}. For all ions up to Dy the $P\overline{3}m$1 structure is realized with considerable DM interaction and a magnetic transition temperature around 15~K with a propagation vector q = 0 for the magnetic structure. In a second magnetic transition at half of T$_N$ the rare-earth ion is polarized  whereas for all superstructure variants crystallizing in a structure described by the space group $R\overline{3}$ we find a magnetic transition around T$_N$~=~2.5~K. For the latter magnetic order is described by an in-plane propagation vector q = (1/3 1/3). The magnetic order remains frustrated and is close to a SL phase, resulting in a largely reduced release of entropy and a 1/3 magnetization plateau even for the ordered systems. Notably in good agreement with the determination of magnetic superexchange interaction we find that with chemical pressure the system distorts more and the couplings thus reduce resulting in a reduction of the magnetic transition temperature T$_N$. Our findings reveal a robust ground state against chemical pressure that can be subtly controlled by the size of the rare-earth ion. The rare-earth ions  remain paramagnetic in vanishing external fields. Structural disorder however, appears to easily destabilize this ground state and due to the vicinity to a SL phase induces this ground state. As our $\mu$SR study reveals no clear LRO for the Yb$_3$Cu$_9$(OH)$_{19}$Br$_{8}$ system we presume that here some finite disorder is present. Likely this arises from the quite different synthesis route, where KBr is used as the halide source and crystallization is based on the slow diffusion out of the teflon liner.

\section*{Summary}
We have shown that phases with composition $A_3$Cu$_9$(OH)$_{19}$Cl$_{8}$ with $A$~=~Nd, ...,Lu  can be  synthesized. The series realizes a chemical pressure tunable kagome lattice, where the rare-earth ion remains paramagnetic until they are exposed to some stronger magnetism either in form of an external field or due to ordering of the Cu$^{2+}$ magnetic moments. Starting from Dy clear superstructure reflexes are observed forming a distorted kagome lattice as a consequence of chemical substitution of Cl$^-$ by OH$^-$, which has a shorter bond and pulls out the $A-$site atom. The distorted kagome has a phase diagram with a large SL phase in the center. All $A_3$Cu$_9$(OH)$_{19}$Cl$_{8}$ are relatively close to the SL phase and negative chemical pressure clearly pushes these systems closer as it leads to large Cu-O-Cu bond angles increasing the superexchange. In addition external pressure control and disorder can bring the system into the SL phase, as the magnetic LRO is fragile, with an uniform reduced moment and low entropy of 10\% in all cases. 
All superstructure variants show a clear 1/3 magnetization plateau that are predicted for the SL ground state. Our detailed characterization on  Yb$_3$Cu$_9$(OH)$_{19}$Br$_{8}$, a system further away from the SL border than the Y counterpart, but the large moment from Yb allows as to push the 1/3 plateau to low fields of 3 T giving an interesting candidate where future neutron studies will shed light on the complex nature of these ground states.

\bibliography{bib}

\section*{Methods}
\subsection*{Sample Preparation and Composition}
Crystals of YCu$_3$(OH)$_{6}$Cl$_{3}$ and Y$_3$Cu$_9$(OH)$_{19}$Cl$_{8}$ were grown by two different routes following the descriptions provided in Refs.~\onlinecite{Sun2016,Biesner2022,Chatterjee2023}: The first route utilized a waterfree hydroflux variant starting from a mixture of 0.6g LiOH, 1.5g BaCl$_2$, 2g Y(NO$_3$)$_3\cdot6$H$_2$O,	1g CuCl$_2\cdot$2H$_2$O, and 0.6g CuCl which was heated to 240$\degree$C for four days in a 23 ml teflon lined autoclave. The second route used hydrothermal transport of CuO in a YCl$_3\cdot$H$_2$O solution at elevated temperatures of 240$\degree$C at the hot end of thick walled Pyrex glass ampoules ($\phi_{inner}$=15 mm, $\phi_{outer}$=21 mm, $l=150$ mm).
Similarly the compounds with the rare earth elements Sm - Eu were grown by hydrothermal transport \cite{Puphal2018}, while Gd - Lu were grown by standard hydrothermal synthesis in a furnace with  a temperature gradient as described in detail in Ref. \cite{Puphal2017}, however with 23 ml teflon lined autoclaves. 
Y$_3$Cu$_9$(OH)$_{19}$Br$_{8}$ crystals of the Br family were synthesized following  Ref.~\cite{Chen2020} with an alternation given in detail in Ref. \cite{Kremer2024}.  Gd$_3$Cu$_9$(OH)$_{19}$Br$_{8}$ was prepared as described in Ref.~\cite{Cheng2022}.  0.6 g Cu(NO$_3$)$_2\cdot$3H$_2$O, 1.9 g Y(NO$_3$)$_3\cdot$6H$_2$O and 2.2 g KBr were mixed and  dissolved in 2 mL deionized water.  Alternatively 0.4514 g Gd(NO$_3)_3\cdot6$H$_2$O  (1 mmol), 0.7248 g Cu(NO$_3$)$_2$ $\cdot3$H$_2$O, 0.7203 g NaBr (7 mmol) were dissolved in 1 ml deionized water. We report the first synthesis of Yb$_3$Cu$_9$(OH)$_{19}$Br$_{8}$. We used a slurry of 0.754 Cu(NO$_3$)$_2$ $\cdot$ 3H$_2$O, 0.7866  Yb(NO$_3$)$_3\cdot5$H$_2$O, 1.459 KBr	and 2ml H$_2$O. Without further homogenization the slurries were transferred to a 23 mL teflon lined autoclave and heated to 230$\degree$C for 3 days.
All Br synthesis differ from that of the Cl samples as crystallization is rather a consequence of over-saturation and less of supercooling. Temperature variations confirmed this observation since higher temperatures led to a faster concentration change and different Br deficient phases were found. We tested variations of the composition, and the temperature profile and noticed that, for both systems, it is important to  apply a temperature above 180$\degree$C as otherwise the parent compound Cu$_2$(OH)$_3$(Br,Cl) form which can be readily detected since e.g. Cu$_2$(OH)$_3$Br shows a magnetic transition at around 10~K \cite{Zheng2009}. Notably, perfect quenching of a hot water solution is difficult and tiny impurities of Cu$_2$(OH)$_3$Br often cannot be avoided.

\subsection*{Single Crystal XRD}
XRD at room temperature was performed on as grown single crystals of typical lateral sizes of 300 $\mu$m. The crystals were attached with vacuum grease on a glass capillary. Diffraction data were collected with a Rigaku miniflex II, using graphite-monochromated  Mo-K$_{\alpha}$ radiation ($\lambda$~=~0.71073~\AA).
The reflection intensities were integrated with the Rigaku software CrysAlis. All the diffraction data show patterns of reverse/obverse twins, typical for rhombohedral symmetry. The crystal structure was solved using Olex2 via the ShelX packages.

\subsection{Energy Dispersive X-ray}
 Energy-dispersive X-ray spectra (EDX) were recorded with a NORAN System 7 (NSS212E) detector in a Tescan Vega SEM (TS-5130MM) on flat hexagonal crystals, which were carbon sputtered.
 \\

\subsection*{Magnetization and Specific Heat}
Magnetization measurements were performed on single crystals at temperatures $0.4\le T \le 350$~K using a superconducting quantum interference device (MPMS-XL, Quantum Design) equipped with the He$^3$ option. 
Specific-heat measurements were carried out, down to $T$ = 400~mK in a {Physical Properties Measurement System} (PPMS, Quantum Design) similarly equipped with the He$^3$ option. 

\subsection*{$\mu$SR Experiments}
$\mu$SR measurements were performed with longitudinal field polarization in the range of 50 down to 0.03~K in zero-field (ZF), weak transverse field (TF), and longitudinal field (LF) on the instrument FLAME installed at the PSI muon facility. A set of  of coaligned phase pure single crystals of sizes $\sim 1\times1\times0.1$~mm$^3$ with their $c$ axis parallel to the initial muon spin direction were used.

\section*{Author contributions statement}
The project was conceived and all experiments and analysis were carried out by P.P. The He$^3$ measurements were supervised by R.K. $\mu$SR experiments were conducted at PSI by J.A.K, T.J.H. H.L. and P.P. The manuscript was written by P.P. All authors reviewed the manuscript. 

\section*{Data availability}
The data are available upon request.

\section*{Competing interests}
The authors declare no competing interests.
\end{document}